\documentclass[twocolumn,showpacs,preprintnumbers,amsmath,amssymb]{revtex4}

\usepackage{graphicx}
\usepackage{bm}

\newtheorem{Lemma}{Lemma}

\newcommand{\Real}{{\textrm{Re}}}

\begin{document}

\title{Quantum adiabatic search with decoherence in the instantaneous
 energy eigenbasis}

\author{Johan {\AA}berg}
\email{johan.aaberg@kvac.uu.se}
\author{David Kult}
\email{david.kult@kvac.uu.se}
\author{Erik Sj\"oqvist}
\email{eriks@kvac.uu.se}
\affiliation{Department of Quantum Chemistry,
 Uppsala University, Box 518, SE-751 20 Uppsala, Sweden}

\begin{abstract}
In Phys. Rev. A {\bf 71}, 060312(R) (2005) the robustness of the
local adiabatic quantum search to decoherence in the instantaneous
eigenbasis of the search Hamiltonian was examined.  We expand this
analysis to include the case of the global adiabatic quantum
search. As in the case of the local search the asymptotic time
complexity for the global search is the same as for the ideal closed
case, as long as the Hamiltonian dynamics is present. In the case of
pure decoherence, where the environment monitors the search
Hamiltonian, we find that the time complexity of the global quantum
adiabatic search scales like $N^{3/2}$, where $N$ is the list
length. We moreover extend the analysis to include success
probabilities $p<1$ and prove bounds on the run time with the same
scaling as in the conditions for the $p\rightarrow 1$ limit.  We
supplement the analytical results by numerical simulations of the
global and local search.
\end{abstract}


\maketitle
\section{\label{sec:intro}Introduction}

Quantum adiabatic algorithms, first conceived in Ref. \cite{Farhi00},
have been developed as an alternative to the traditional circuit model
of quantum computation. An important aspect of the adiabatic approach
is its expected resilience to various kinds of open system effects,
which makes it a promising candidate for robust quantum
computation. The basic reason for this appealing feature is that the
adiabatic quantum computer operates near the instantaneous ground
state of a time-dependent Hamiltonian and can therefore be expected to
be insensitive to relaxation and open system effects among the excited
states. The robustness of adiabatic quantum computation has been
discussed in the literature for different kinds of perturbations, such
as relaxation \cite{Science}, noise
\cite{noise}, decoherence \cite{Science,scale,aberg05}, and unitary 
control errors \cite{robust}. 

In this paper, we elaborate on the robustness of the quantum adiabatic
search \cite{Farhi00,Roland02,vanDam01} in the presence of decoherence
in the instantaneous eigenbasis of the search Hamiltonian. In
particular, we extend the analysis of the local search
\cite{Roland02,vanDam01} in Ref.~\cite{aberg05} 
to global adiabatic search \cite{Farhi00}. We further 
address the dependence of the asymptotic time complexity of the search
procedure on the success probability for finding the searched item.

To model adiabatic search in the presence of decoherence we consider 
master equations of the form ($\hbar = 1$ from now on)
\begin{equation}
\label{enkel}
\dot{\varrho} = -iA[H,\varrho] -B\bm{[}W,[W,\varrho]\bm{]},
\end{equation}
where $\varrho$ is the density operator, and $A,B$ are constants such
that $B\geq 0$. For convenience, we also assume that $A\geq 0$. The
operator $H$ is taken to be the time-dependent Hamiltonian of the
adiabatic search \cite{Farhi00,Roland02}, and $W$ is a time-dependent
Hermitian operator such that $[H,W] = 0$, which gives decoherence in
the instantaneous eigenbasis of $H$.

Decoherence with respect to the energy eigenbasis of a system has been
shown to occur in certain regimes of system-environment interaction
\cite{Zurek}.  Apart from this general regime of weak interaction and
dominating self-Hamiltonians one can find other settings where this
type of decoherence occurs. One example is control errors in terms of
fluctuations in the energy levels of the system \cite{note}.  Note
also that decoherence in the energy eigenbasis does not necessarily
imply the particular form of Eq.~(\ref{enkel}). However, this choice
provides a sufficiently simple model to be treated analytically, and
yet results in non-trivial behavior. Moreover, the form of
Eq.~(\ref{enkel}) quite generally guarantees the solution to be a
proper density operator at all times.

Our results show that for $A\neq 0$ and list length $N$, it is
sufficient with a run time of the order $N$ in the global case and
$\sqrt{N}$ in the local case, as is the case for the ideal closed
evolution. These results suggest that the adiabatic quantum computer
is protected against this type of decoherence, which is promising for
physical realizations. In the wide-open case where $A =0$
\cite{percival}, with the special choice $W = H$, the scaling changes to
$N^{3/2}$ and $N$, for the global and local search, respectively.

Although the concept of adiabaticity has a well defined meaning for
closed systems, there is no obviously unique generalization to the
case of open systems. One may consider several different
generalizations focusing on various aspects of the standard
adiabaticity. For example, in Refs.~\cite{Sarandy04,SaWuLi,sarandy05}
the concept of adiabaticity is based upon Jordan block decompositions
of the superoperator that describe the evolution of the open system. A
different approach, designed for systems that are weakly open, has
been put forward in Ref.~\cite{thunstrom05}. This latter approach is
based upon taking decoupling of the energy subspaces of the ideal
Hamiltonian as the starting point. Here, we use a third method based
upon a feature of the adiabatic quantum computer, namely that the
evolution takes place near an eigenstate of the ideal
Hamiltonian. This property is in contrast with, e.g., implementations
of holonomic quantum gates
\cite{holonomic}, where it is essential that the gate can operate on
arbitrary superpositions, without too large errors. For the
functioning of the adiabatic quantum computer, however, it is
sufficient that the state of the system remains close to the correct
eigenstate. In the present interpretation of the word ``adiabaticity''
we only require that the probabilities of finding the system in the
instantaneous eigenstates of $H(s)$ are conserved, but we allow
superpositions of the instantaneous eigenstates to decay into
mixtures.

The structure of the article is as follows. In Sec.~\ref{sec:twolev}
we derive an integral equation from the original master
equation. Section \ref{sec:search} gives a brief overview of the
adiabatic search algorithm, and we calculate some quantities which are
used later. In Sec.~\ref{sec:semiopen} we prove a sufficient condition
for the success probability of the adiabatic search to approach unity
in the semi-open case.  We consider both global and local
search. Furthermore, a condition on the decoherence term of the master
equation, which is used in the proofs in Sec.~\ref{sec:semiopen}, is
examined.  In Sec.~\ref{sec:wopen} we demonstrate that the wide-open
case is essentially different from the semi-open case.  We prove
sufficient conditions for the success probability to approach unity in
both the global, as well as the local case. We further prove that the
sufficient conditions are also necessary.  In Sec.~\ref{sec:success},
bounds on the run time for fixed success probabilities less that
unity, are given.  We end with the conclusions.

\section{\label{sec:twolev}The master equation}
In this section we develop the basic equation used in our
analysis. Since it is our intention to use adiabaticity we consider a fixed
family of Hamiltonians $H(s)$, parametrized, in a sufficiently smooth
way, by the parameter $s$. We assume that the parameter $s$ is
proportional to the time $t$, such that $s = 0$ at time $t=0$ and
$s=1$ at time $t = T$. The time $T$ is referred to as the ``run
time''. We also assume a family of Hermitian operators $W(s)$, to be
used in the master equation
\begin{equation}
\label{startekv}
\frac{d}{dt}\varrho(t) =
 -iA[H(t/T),\varrho(t)] -B\bm{[}W(t/T),[W(t/T),\varrho(t)]\bm{]}.
\end{equation}
We assume that both $H(t/T)$ and $W(t/T)$ are Hermitian and
non-degenerate at each $t$. Moreover, we assume $[H(t/T),W(t/T)] = 0$
for all $t$.  By a change of variables $s = t/T$ in Eq.~(\ref{startekv})
one obtains
\begin{equation}
\label{sparam}
\frac{d}{ds}\rho(s) =  -iTA[H(s),\rho(s)] -TB\bm{[}W(s),[W(s),\rho(s)]\bm{]},
\end{equation}
where $\rho(s) = \varrho(sT)$.  Let $\{|E_{n}(s)\rangle\}_{n}$ be
instantaneous orthonormal eigenvectors of $H(s)$, i.e.,
$H(s)|E_{n}(s)\rangle = E_{n}(s)|E_{n}(s)\rangle$. Since $H(s)$ is
assumed to be non-degenerate the arbitrariness in the choice of
eigenvectors is limited to phase factors. Let $w_{n}(s)$ denote the
eigenvalues of $W(s)$.  If we insert $\rho(s) =
\sum_{nm}\rho_{nm}(s)|E_{n}(s)\rangle\langle E_{m}(s)|$
into Eq.~(\ref{sparam}), the equations for the components
$\rho_{nm}(s)$ become
\begin{eqnarray}
\label{grundekv}
\frac{d}{ds}\rho_{nm}(s) & = & -iTA\{ E_{n}(s)-E_{m}(s)\}\rho_{nm}(s)
\nonumber\\*
 & & -TB\{w_{n}(s)-w_{m}(s)\}^{2}\rho_{nm}(s)\nonumber\\* & &
 -i[\bm{Z}(s),\bm{\rho}(s)]_{nm},
\end{eqnarray}
where $\bm{Z}(s)$ is an Hermitian matrix with elements
\begin{equation}
\label{Zdef}
Z_{nm}(s) = i\langle \dot{E}_{n}(s)|E_{m}(s)\rangle
\end{equation}
and $\bm{\rho}(s)$ is the matrix with elements
$\rho_{nm}(s)$.

Consider an ordinary differential equation
\begin{equation}
\label{sftjf}
\frac{d}{ds}x(s) = L_{1}(s)x(s) + L_{2}(s)x(s), 
\end{equation}
where $L_{1}(s)$ and $L_{2}(s)$ are s-dependent linear operators
acting on the vector $x$. If the space is finite-dimensional,
$L_{1}(s)$ and $L_{2}(s)$ are sufficiently well behaved as functions
of $s$, and if $L_{1}(s)$ fulfills $[L_{1}(s),L_{1}(s')] = 0$ for all
$s$, $s'$, then Eq.~(\ref{sftjf}) can be rewritten as the following
integral equation
\begin{eqnarray}
x(s) & = & e^{\int_{0}^{s} L_{1}(s')ds'}x(0)\nonumber\\
& & + e^{\int_{0}^{s}L_{1}(s')ds'}
\int_{0}^{s}e^{-\int_{0}^{s'}L_{1}(s'')ds''}L_{2}(s')x(s')ds'.\nonumber\\
\end{eqnarray}
In the present case we have $x(s) = \bm{\rho}(s)$ and 
\begin{eqnarray}
(L_{1}(s)\bm{\rho}(s))_{nm} & = & -
iTA\{E_{n}(s)-E_{m}(s)\}\rho_{nm}(s)\nonumber\\ & & -TB
\{w_{n}(s)-w_{m}(s)\}^{2} \rho_{nm}(s),\\ L_{2}(s)\bm{\rho}(s) & = &
-i[\bm{Z}(s),\bm{\rho}(s)].
\end{eqnarray} 
Hence, Eq.~(\ref{grundekv}) can be rewritten as
\begin{eqnarray}
\label{grundint}
\rho_{nm}(s) &=& e^{-iTAR_{nm}(s)-TBQ_{nm}(s)}\rho_{nm}(0)\nonumber\\
& & -ie^{-iTAR_{nm}(s)-TBQ_{nm}(s)}\nonumber \\ &
&\times\int_{0}^{s}e^{iTAR_{nm}(s')
+TBQ_{nm}(s')}[\bm{Z}(s'),\bm{\rho}(s')]_{nm}ds',\nonumber\\
\end{eqnarray}
where 
\begin{equation}
Q_{nm}(s) = \int_{0}^{s}[w_{n}(s')-w_{m}(s')]^{2}ds',
\end{equation}
\begin{equation}
R_{nm}(s) = \int_{0}^{s}[E_{n}(s')-E_{m}(s')]ds'.
\end{equation}

\section{\label{sec:search}The search problem}
Our task is to find a single marked item in a collection of $N$ items.
The $N$-element search problem is associated with an $N$-dimensional
Hilbert space with orthonormal basis $\{|k\rangle\}_{k=1}^N$, where
the marked item corresponds to $|\mu\rangle \in
\{|k\rangle\}_{k=1}^N$.
We consider the following family of Hamiltonians \cite{Farhi00,Roland02},
\begin{equation}
\label{family}
H(t/T) = -(1-t/T)|\psi\rangle\langle\psi| - t/T|\mu\rangle\langle
\mu|,
\end{equation}
where
\begin{equation}
\label{eqalsuperpos}
|\psi\rangle = \frac{1}{\sqrt{N}}\sum_{k=1}^N |k\rangle.
\end{equation}
The ground state $|\psi\rangle$ of the initial Hamiltonian $H(0)$ is
easy to prepare, while the ground state of $H(1)$ gives the solution
of the search problem. Hence, if we assume the evolution is adiabatic,
this family of Hamiltonians takes us from the initial state
$|\psi\rangle$ to the marked state $|\mu\rangle$, and thus solves the
search problem.  A crucial observation is that the problem is
essentially two-dimensional, since the space ${\cal H}_R$ spanned by
$|\psi\rangle$ and $|\mu\rangle$ is the only relevant subspace. We may
Gram-Schmidt orthogonalize $\{|\psi\rangle,|\mu\rangle\}$, yielding
the orthonormal basis
\begin{equation}
\label{basis}
\{|\psi\rangle,|\overline{\psi}\rangle\}, \quad |\overline{\psi}\rangle
= \frac{\sqrt{N}|\mu\rangle-|\psi\rangle}{\sqrt{N-1}}.
\end{equation}
We represent the Hamiltonian family in Eq.~(\ref{family}), restricted
to the space ${\cal H}_R$, as matrices in the basis defined in
Eq.~(\ref{basis}). After the substitution $s=t/T$, we get
\begin{equation}
\label{Mmatris}
\bm{M}(s) = \left[\begin{matrix} s\frac{N-1}{N}-1& -s\frac{\sqrt{N-1}}{N}\\
-s\frac{\sqrt{N-1}}{N} & -s\frac{N-1}{N}
\end{matrix}\right],
\end{equation}
which has the eigenvalues
\begin{equation}
\label{egenvarden}
E_{0}(s) = -\frac{1}{2}- \frac{1}{2}\Delta(s),\quad E_{1}(s) =
-\frac{1}{2}+ \frac{1}{2}\Delta(s)
\end{equation}
and orthonormal eigenvectors
\begin{eqnarray}
\label{sdfh}
\bm{e}^{(0)} & = & \frac{\sqrt{2}}{\sqrt{\Delta^{2} -\Delta D}}
\left[\begin{matrix} \frac{1}{2}(\Delta-D)\\ s\frac{\sqrt{N-1}}{N}
\end{matrix}\right]
 = \left[\begin{matrix}e^{(0)}_{1}(s) \\ e^{(0)}_{2}(s)\end{matrix}\right],
\nonumber\\
\bm{e}^{(1)} & = & \frac{\sqrt{2}}{\sqrt{\Delta^{2}  +\Delta D}}
\left[\begin{matrix}
\frac{1}{2}(\Delta + D)\\ -s\frac{\sqrt{N-1}}{N}
\end{matrix}\right] = 
\left[\begin{matrix} e^{(1)}_{1}(s)\\ e^{(1)}_{2}(s)\end{matrix}\right],
\end{eqnarray}
where
\begin{eqnarray}
\label{Deltadef} \Delta(s) & = & E_{1}(s)-E_{0}(s)\nonumber\\
& =&\sqrt{ \frac{1+(N-1)(2s-1)^{2}}{N}},\\
\label{Ddef} D(s) & = & -1+2s\frac{N-1}{N}.
\end{eqnarray}
The two relevant eigenvectors of $H(s)$ are
\begin{eqnarray}
\label{vektorer}
|E_{0}(s)\rangle & = & e^{(0)}_{1}(s)|\psi\rangle +
e^{(0)}_{2}(s)|\overline{\psi}\rangle,\nonumber\\
|E_{1}(s)\rangle & = & e^{(1)}_{1}(s)|\psi\rangle +
e^{(1)}_{2}(s)|\overline{\psi}\rangle,
\end{eqnarray}
where $|E_{0}(s)\rangle$ is the ground state.

We next calculate the matrix $\bm{Z}(s)$, defined in
Eq.~(\ref{Zdef}). In order to save some efforts we first notice that
$Z_{00}(s)$ as well as $Z_{11}(s)$ are identically zero, since
$e^{(0)}_{0}$, $e^{(0)}_{1}$, $e^{(1)}_{0}$, and $e^{(1)}_{1}$ are
real valued. Secondly $\bm{Z}(s)$ has to be an Hermitian matrix, which
follows from differentiation of the relation $\langle
E_{n}(s)|E_{m}(s)\rangle =
\delta_{nm}$. 

By combining Eq.~(\ref{Zdef}) with Eqs.~(\ref{sdfh}),
(\ref{Deltadef}), and (\ref{Ddef}), one obtains the following result
\begin{equation}
\label{Z01}
Z_{01}(s) =  \frac{-i\sqrt{N-1}}{1+ (N-1)\left(2s-1\right)^{2}}. 
\end{equation}
Define 
\begin{equation}
Z(s) = |Z_{01}(s)| = \frac{\sqrt{N-1}}{1+(N-1)(2s-1)^{2}}.
\end{equation}
Note that the
maximum $\sqrt{N-1}$ of $Z(s)$ is obtained at $s = 1/2$. A useful
property of $Z$ is
\begin{equation}
\label{Zegensk}
\int_{0}^{1}Z(s)ds \leq\frac{\pi}{2},
\end{equation}
for all $N\geq 1$.

Concerning the reduction to the two-dimensional problem, one
should note that, due to the assumption $[H(s),W(s)] =0$, the operator
$W(s)$ block-diagonalizes with respect to the eigenspaces of
$H(s)$. This implies that the solution of Eq.~(\ref{startekv})
remains in the subspace ${\cal H}_R$ once
started there. Thus, we only need to consider the master equation in
Eq.~(\ref{startekv}) on this particular subspace.

Define
\begin{eqnarray}
\label{QRdef}Q(s) & =&  Q_{10}(s),\quad R(s) = R_{10}(s),\\
\label{Ydef}Y(s) & =& \rho_{00}(s)-\rho_{11}(s) = 2\rho_{00}(s)-1.
\end{eqnarray}
From Eq.~(\ref{grundint}) one can derive the following  equation for $Y(s)$,
\begin{widetext}
\begin{eqnarray}
\label{tvaanivaa}
Y(s) &=& Y(0) -
4\int_{0}^{s}e^{-TBQ(s')}Z(s')\Real\left[e^{-iTAR(s')}\rho_{10}(0)\right]ds'
\nonumber\\*
& & - 4 \int_{0}^{s} e^{-TBQ(s')}\int_{0}^{s'}e^{TBQ(s'')}
\Real\left[e^{-iTAR(s')}e^{iTAR(s'')}\right]Z(s')Z(s'')Y(s'')ds''ds'.
\end{eqnarray}
\end{widetext}
As seen from Eq.~(\ref{Ydef}), the value of $Y(s)$ determines to what
degree the system is in the ground state, i.e., $Y=1$ when the system
is in the ground state, and $Y =-1$ when the system is in the first
excited state.

\section{\label{sec:semiopen}Semi-open system}
\subsection{\label{sec:semiopenglobal}Global search}
We begin with a comment on our terminology regarding the ``openness''
of quantum systems. With a ``wide-open'' \cite{percival} system we
intend a system governed by Eq.~(\ref{enkel}), with $A =0$. As a
contrast to the wide-open case we define ``semi-open'' as
$A>0$. Finally, a ``closed'' system denotes the case $B = 0$.

We define
\begin{eqnarray}
\Gamma(s) & = & w_{1}(s)-w_{0}(s),\\
R(s) & = & \int_{0}^{s}\Delta(s')ds',\\ Q(s) & = &
\int_{0}^{s}\Gamma^{2}(s')ds',\\
\label{lideltadef} \delta &=& \min_{s\in [0,1]}\Delta(s) =
 \frac{1}{\sqrt{N}},\\
\label{ligammadef} \gamma & = & \min_{s\in [0,1]}\Gamma(s).
\end{eqnarray}
In the proofs we use the following restriction on the decoherence term
\begin{equation}
\label{condition}
\int_{0}^{1}Z(s)\left|\frac{d}{ds}\Gamma^{2}(s)\right| ds \leq K,
\end{equation}
where $K$ is a constant independent of $N$.  This assumption
essentially means that the fluctuations of $\Gamma$ in $s$ are not
allowed to grow with the list-length $N$. The restriction to an
$N$-independent $K$ can be relaxed. We discuss this issue, as well as
the condition in general, in Sec.~\ref{sec:condition}.
 
From Eq.~(\ref{tvaanivaa}) it follows that
\begin{equation}
\label{diff}
 |Y(0)-Y(s)| \leq  4 \left|J(s)\right| + 4\left|I(s)\right|,
\end{equation}
where 
\begin{eqnarray}
\label{Iuppd} I(s) & = & \frac{1}{2}I_{+}(s) + \frac{1}{2}I_{-}(s),\\
\label{Ipmdef} I_{\pm}(s) & = 
&\int_{0}^{s} e^{-T[BQ(s')\pm iAR(s')]}Z(s')u_{\pm}(s')ds',\\
\label{gIdef} u_{\pm}(s') & = 
& \int_{0}^{s'}e^{T[BQ(s'')\pm iAR(s'')]}Z(s'')Y(s'')ds'',\nonumber\\
\end{eqnarray}
and
\begin{equation}
\label{Jdef}
J(s) =
\int_{0}^{s}e^{-TBQ(s')}Z(s')\Real\left[e^{-iTAR(s')}\rho_{10}(0)\right]ds'.
\end{equation}
For the sake of simplicity we separate the treatment of the
terms $I$ and $J$.

\subsubsection{The term $I$}
To begin with we define the following integral and perform a partial
integration
\begin{widetext}
\begin{eqnarray}
\label{Gdef}
G_{\pm}[g](s) & = & \int_{0}^{s}e^{- T[BQ(s')\pm
iAR(s')]}Z(s')g(s')ds'\nonumber\\ & = & \int_{0}^{s}e^{- T[BQ(s')\pm
iAR(s')]}\left[B\frac{dQ}{ds'}(s')\pm
iA\frac{dR}{ds'}(s')\right]\frac{Z(s')g(s')}{B\frac{dQ}{ds'}(s')\pm
iA\frac{dR}{ds'}(s')}ds'\nonumber\\* &= & -\frac{1}{T}e^{- T[BQ(s)\pm
iAR(s)]}\frac{Z(s)g(s)}{B\frac{dQ}{ds}(s)\pm
iA\frac{dR}{ds}(s)}\nonumber\\ &
&+\frac{1}{T}\frac{\sqrt{N-1}}{N}\frac{g(s)}{B\frac{dQ}{ds}(s)\pm
iA\frac{dR}{ds}(s)}\bigg|_{s = 0}\nonumber\\* & &
+\frac{1}{T}\int_{0}^{s}e^{- T[BQ(s')\pm
iAR(s')]}\frac{d}{ds'}\left(\frac{Z(s')}{B\frac{dQ}{ds'}(s')\pm
iA\frac{dR}{ds'}(s')}\right)g(s')ds'\nonumber\\* & &
+\frac{1}{T}\int_{0}^{s}e^{- T[BQ(s')\pm
iAR(s')]}\frac{Z(s')}{B\frac{dQ}{ds'}(s')\pm
iA\frac{dR}{ds'}(s')}\frac{d}{ds'}g(s')ds'.
\end{eqnarray}
If we put $g(s) =  u_{\pm}(s)$ into Eq.~(\ref{Gdef}) the result is
\begin{eqnarray}
\label{Ipmutr}
I_{\pm}(s) & = & G_{\pm}[u_{\pm}](s) \nonumber\\
& = & -\frac{1}{T}e^{-T[BQ(s)\pm
iAR(s)]}\frac{Z(s)}{B\frac{dQ}{ds}(s)\pm iA\frac{dR}{ds}(s)}
\int_{0}^{s}e^{T[BQ(s')\pm
iAR(s')]}Z(s')Y(s')ds'\nonumber\\* 
& & +\frac{1}{T}\int_{0}^{s}e^{-T[BQ(s')\pm iAR(s')]}
\frac{d}{ds'}\left(\frac{Z(s')}{B\frac{dQ}{ds'}(s')\pm
iA\frac{dR}{ds'}(s')}\right)\int_{0}^{s'}e^{T[BQ(s'')\pm
iAR(s'')]}Z(s'')Y(s'')ds'' ds'\nonumber\\* 
& &+\frac{1}{T}\int_{0}^{s}\frac{Z(s')}{B\frac{dQ}{ds'}(s')\pm
iA\frac{dR}{ds'}(s')}Z(s')Y(s')ds'.
\end{eqnarray}
\end{widetext}
In the next step we calculate an upper bound to $|I_{\pm}(s)|$.  To do
this we use that $|Y(s)|\leq 1$. Moreover, since $s'\geq s''$ it
follows that $Q(s')\geq Q(s'')$, which implies
$\exp\{-TB[Q(s')-Q(s'')]\}\leq1$. We also use that $dQ/ds =
\Gamma^{2}(s)$, $dR/ds = \Delta(s)$, and Eq.~(\ref{Zegensk}) to obtain
\begin{eqnarray}
|I_{\pm}(s)| & \leq &\frac{1}{T}\frac{\pi}{2}\int_{0}^{s}
\left|\frac{d}{ds'}\left(\frac{Z(s')}{B\Gamma^{2}(s')\pm
iA\Delta(s')}\right)\right| ds'\nonumber\\
& &+\frac{1}{T}\int_{0}^{s}\frac{Z^{2}(s')}{\sqrt{B^{2}\Gamma^{4}(s')+
A^{2}\Delta^{2}(s')}}ds'\nonumber\\
& & +\frac{1}{T}\frac{\pi}{2}\frac{Z(s)}{\sqrt{B^{2}\Gamma^{4}(s)+
A^{2}\Delta^{2}(s)}}.
\end{eqnarray}
Now we use the fact that $Z(s)\leq \sqrt{N-1}$, as well as
Eqs.~(\ref{lideltadef}) and (\ref{ligammadef}), which result in
\begin{eqnarray}
\label{Idelsteg}
|I_{\pm}(s)| &\leq &\frac{1}{T}\frac{\pi}{2}\int_{0}^{s}
\left|\frac{d}{ds'}\left(\frac{Z(s')}{B\Gamma^{2}(s')\pm
iA\Delta(s')}\right)\right| ds'\nonumber\\
& & +\frac{\pi}{T}\frac{\sqrt{N-1}}{\sqrt{B^{2}\gamma^{4}+
A^{2}\delta^{2}}}.
\end{eqnarray}
To treat the integrand in Eq.~(\ref{Idelsteg}), we define
\begin{equation}
X(s) = \left|\frac{d}{ds}\left(\frac{Z(s)}{B\Gamma^{2}(s)\pm
iA\Delta(s)}\right)\right|, \nonumber\\
\end{equation}
which can be estimated as
\begin{eqnarray}
\label{Xolikhet}
X(s) &\leq & \frac{1}{\sqrt{B^{2}\gamma^{4}+ A^{2}\delta^{2}}}
\left|\frac{d}{ds}Z(s)\right|
\nonumber\\* 
& & + \frac{B}{B^{2}\gamma^{4}+
A^{2}\delta^{2}}Z(s)\left|\frac{d}{ds}\Gamma^{2}(s)\right| \nonumber\\
& & + \frac{A}{B^{2}\gamma^{4}+
A^{2}\delta^{2}}Z(s)\left|\frac{d}{ds}\Delta(s)\right|.
\end{eqnarray}
The function $Z(s)$ is symmetric around $s = 1/2$ and  
increasing on the interval $[0,1/2]$. It follows that the function
$|dZ/ds|$ is also symmetric around $s = 1/2$, and that $|dZ/ds| = dZ/ds$
on $[0,1/2]$. Thus
\begin{equation}
\label{absderint}
 \int_{0}^{s} \left|\frac{d}{ds'}Z(s')\right| ds'\leq \int_{0}^{1}
 \left|\frac{d}{ds'}Z(s')\right| ds'\leq 2\sqrt{N-1}.
\end{equation}
Since both $Z(s)$ and $\Delta(s)$ are symmetric around $s = 1/2$,
$Z(s)|d\Delta/ds|$ has the same symmetry. Moreover, $\Delta(s)$ is
increasing on the interval $[1/2,1]$. Hence, $Z(s)|d\Delta/ds|
=Z(s)d\Delta/ds$ on $[1/2,1]$, which leads to
\begin{eqnarray}
\label{ZderD}
\int_{0}^{s}Z(s)\left|\frac{d}{ds}\Delta(s)\right|ds 
& \leq & \int_{0}^{1}Z(s)\left|\frac{d}{ds}\Delta(s)\right|ds\nonumber\\ 
& \leq & 2\sqrt{\frac{N-1}{N}} \leq 2.
\end{eqnarray}
By combining Eq.~(\ref{condition}) with Eqs.~(\ref{Xolikhet}),
(\ref{absderint}), and (\ref{ZderD}), one obtains
\begin{equation}
\label{absd}
\int_{0}^{s}X(s')ds' \leq \frac{2\sqrt{N-1}}{\sqrt{B^{2}\gamma^{4}+
 A^{2}\delta^{2}}} +  \frac{BK+2A}{B^{2}\gamma^{4}+ A^{2}\delta^{2}}.
\end{equation}
By combining Eqs.~(\ref{lideltadef}), (\ref{Iuppd}), (\ref{Idelsteg}),
and (\ref{absd}) it follows that
\begin{equation}
\label{Iresultat}
|I(s)| \leq
\frac{\pi}{2}\frac{N}{T}\left(\frac{4}{\sqrt{B^{2}\gamma^{4}N+ A^{2}}}
+ \frac{BK+2A}{B^{2}\gamma^{4}N+ A^{2}}\right).
\end{equation}

\subsubsection{The term $J$}
Equation (\ref{Jdef}) can be rewritten as
\begin{equation}
\label{Jekv}
J(s) =   
\Real\left(\rho_{10}(0)\int_{0}^{s}e^{-T[BQ(s')+iAR(s')]}Z(s')ds' \right). 
\end{equation}
We write $|J(s)| \leq |\rho_{10}(0)|\,\,|G_{+}[1](s)|$, with $G$ as in
Eq.~(\ref{Gdef}) with $g(s)\equiv 1$.  By a line of reasoning very
similar to the one in the previous section one obtains
\begin{eqnarray}
|G_{+}[1](s)| &\leq &
\frac{1}{T}\int_{0}^{s}\left|\frac{d}{ds'}\left(\frac{Z(s')}{B\Gamma^{2}(s')+
iA\Delta(s')}\right)\right|ds'\nonumber\\
& &+\frac{1}{T\sqrt{N}}\frac{1}{\sqrt{B^{2}\Gamma^{4}(0)+A^{2}}}
\nonumber\\
& &+\frac{1}{T}\frac{Z(s)}{\sqrt{B^{2}\Gamma^{4}(s)+
A^{2}\Delta^{2}(s)}},
\end{eqnarray}
which  results in
\begin{eqnarray}
\label{Jresultat}
|J(s)| &\leq & |\rho_{10}(0)|
\frac{1}{T\sqrt{N}}\frac{1}{\sqrt{B^{2}\Gamma^{4}(0)+A^{2}}}
\nonumber\\ & &+|\rho_{10}(0)|
\frac{N}{T}\frac{3}{\sqrt{B^{2}\gamma^{4}N+ A^{2}}} \nonumber\\ &
&+|\rho_{10}(0)| \frac{N}{T}\frac{BK+2A}{B^{2}\gamma^{4}N+ A^{2}}.
\end{eqnarray}

\subsubsection{Collecting the results}
By combining Eqs.~(\ref{diff}), (\ref{Iresultat}), and (\ref{Jresultat}),
one obtains
\begin{eqnarray}
|Y(0)-Y(s)| &\leq & 4
|\rho_{10}(0)|\frac{1}{T\sqrt{N}}\frac{1}{\sqrt{B^{2}\Gamma^{4}(0)+A^{2}}}
\nonumber\\
& &+ 4 |\rho_{10}(0)| \frac{N}{T}\frac{3}{\sqrt{B^{2}\gamma^{4}N+
A^{2}}}\nonumber\\ & &+ 4
|\rho_{10}(0)|\frac{N}{T}\frac{BK+2A}{B^{2}\gamma^{4}N+ A^{2}}
\nonumber\\ 
& &+8\pi\frac{N}{T}\frac{1}{\sqrt{B^{2}\gamma^{4}N+
A^{2}}}\nonumber\\ 
& &+2\pi\frac{N}{T}
\frac{BK+2A}{B^{2}\gamma^{4}N+ A^{2}}.
\end{eqnarray}
Note that 
\begin{eqnarray}
B^{2}\gamma^{4}N+ A^{2} & \geq  & A^{2},\\
B^{2}\Gamma^{4}(0)+A^{2} & \geq &  A^{2},
\end{eqnarray}
which, as long as $A>0$, implies that
\begin{eqnarray}
\label{sogl}
|\rho_{00}(0)-\rho_{00}(s)| &\leq & 2 |\rho_{10}(0)|
\frac{1}{T\sqrt{N}}\frac{1}{A} \nonumber\\ & & + 2
|\rho_{10}(0)|\frac{N}{T}\frac{BK+5A}{A^{2}}\nonumber \\ & &
+\pi\frac{N}{T}\frac{BK+6A}{A^{2}},
\end{eqnarray}
where we have used Eq.~(\ref{Ydef}).  The leading term in
Eq.~(\ref{sogl}) is proportional to $N/T$.  In the
$N/T\rightarrow\infty$ limit the probability to find the system in the
groundstate is conserved.  Thus, it is a sufficient condition for
adiabaticity that $T \gg N$.

\subsection{\label{sec:semiopenlocal}Local search}
As is well known \cite{Farhi00}, the standard global adiabatic search
algorithm performs as the classical search, but the local adiabatic search
\cite{Roland02,vanDam01} outperforms the classical algorithms. Here we
further elaborate and extend the analysis of Ref.~\cite{aberg05}
concerning the effect of decoherence on the local adiabatic search.
The idea behind the local search is to adjust the speed of parameter
change in such a way that more time is spent near the minimum energy
gap. This is obtained by a reparametrization of the functions $H(s)$
and $W(s)$ to $H\bm{(}f(s)\bm{)}$ and $W\bm{(}f(s)\bm{)}$, where $f$
is a strictly increasing, sufficiently smooth function, such that
$f(0) = 0$ and $f(1) = 1$. The new master equation becomes
\begin{eqnarray}
\label{startlocal}
\frac{d}{ds}\rho(s) & =&
 -iAT[H \bm{(} f(s) \bm{)} ,\rho(s)]\nonumber\\
& & -BT \bm{[} W \bm{(} f(s) \bm{)} ,[W \bm{(} f(s) \bm{)} ,\rho(s)] \bm{]}.
\end{eqnarray}
By a change of variables
$r = f(s)$ and by defining $\widetilde{\rho}(r) = 
\rho \bm{(} f^{-1}(r) \bm{)}$, Eq.~(\ref{startlocal}) becomes
\begin{eqnarray}
\frac{d}{dr}\widetilde{\rho}(r) & =&
 -iAT\frac{df^{-1}}{dr}(r)[H(r),\widetilde{\rho}(r)]\nonumber\\
& & -BT\frac{df^{-1}}{dr}(r)\bm{[}W(r),[W(r),\widetilde{\rho}(r)]\bm{]}.
\end{eqnarray}
Note that $\widetilde{\rho}(0) = \rho(0)$ and $\widetilde{\rho}(1) =
\rho(1)$.

The equation for $\widetilde{Y}(r) = 2\widetilde{\rho}_{00}(r)-1$ is
fully analogous to Eq.~(\ref{tvaanivaa}) with $Q$ and $R$ replaced by
$\widetilde{Q}$ and $\widetilde{R}$, where
\begin{eqnarray}
\widetilde{Q}(r) & = & \int_{0}^{r}\Gamma^{2}(r')\frac{df^{-1}}{dr}dr',\\
\widetilde{R}(r) & = & \int_{0}^{r}\Delta(r')\frac{df^{-1}}{dr}dr'.
\end{eqnarray}
The choice
\begin{eqnarray}
\label{fdefinition}f^{-1}(r) & = & 
\frac{1}{L}\int_{0}^{r}\frac{1}{\Delta^{2}(r')}dr',\\
L &= & \int_{0}^{1}\frac{1}{\Delta^{2}(r')}dr'
\end{eqnarray}
gives the optimal
efficiency for the quantum search problem \cite{Roland02,vanDam01,Grover}.
 One obtains
\begin{eqnarray}
\widetilde{Q}(r) & = & 
\frac{1}{L}\int_{0}^{r}\frac{\Gamma^{2}(r')}{\Delta^{2}(r')}dr',\\
\widetilde{R}(r) & = & \frac{1}{L}\int_{0}^{r}\frac{1}{\Delta(r')}dr',
\end{eqnarray}
\begin{equation}
\label{Ldef} L  =  \frac{N}{\sqrt{N-1}}\arctan(\sqrt{N-1})\leq
\frac{\pi}{2}\frac{N}{\sqrt{N-1}}.
\end{equation}
Define
\begin{equation}
\label{zetadef}
\zeta = \min_{s\in [0,1]}\frac{\Gamma^{2}(s)}{\Delta(s)}.
\end{equation}
As in the global case we let $|\widetilde{Y}(0)-\widetilde{Y}(r)| \leq
4 |\widetilde{J}(r)| + 4|\widetilde{I}(r)|$, and as before we separate
the treatment of the terms $\widetilde{I}$ and $\widetilde{J}$. The
terms $\widetilde{I}(r)$ and $\widetilde{J}(r)$ are defined as in
Eqs.~(\ref{Iuppd})-(\ref{gIdef}) and (\ref{Jekv}), but with $Q$ and
$R$ replaced by $\widetilde{Q}$ and $\widetilde{R}$, respectively.

\subsubsection{The term $\widetilde{I}$}
The analogue of Eq.~(\ref{Ipmutr}), is obtained by replacing $Q$ and
$R$ by $\widetilde{Q}$ and $\widetilde{R}$, respectively. We use
$d\widetilde{Q}/dr = \Gamma^{2}(r)/[L\Delta^{2}(r)]$,
$d\widetilde{R}/dr = 1/[L\Delta(r)]$, 
$\exp\{-TB[\widetilde{Q}(r)-\widetilde{Q}(r')]\}\leq 1$ if $r\geq r'$,
as well as Eq.~(\ref{Zegensk}), which result in
\begin{eqnarray}
|\widetilde{I}_{\pm}(r)| & \leq &
\frac{1}{T}\frac{\pi}{2}\int_{0}^{r}\left|
\frac{d}{dr'}\left(\frac{Z(r')}{B\frac{1}{L}
\frac{\Gamma^{2}(r')}{\Delta^{2}(r')}
\pm iA\frac{1}{L}\frac{1}{\Delta(r')}}\right)\right| dr'\nonumber\\ 
& &+\frac{1}{T}\int_{0}^{r}\frac{Z^{2}(r')}{\sqrt{B^{2}\frac{1}{L^{2}}
\frac{\Gamma^{4}(r')}{\Delta^{4}(r')}+
A^{2}\frac{1}{L^{2}}\frac{1}{\Delta^{2}(r')}}}dr'\nonumber\\ 
& & +\frac{1}{T}\frac{\pi}{2}
\frac{Z(r)}{\sqrt{B^{2}\frac{1}{L^{2}}\frac{\Gamma^{4}(r)}{\Delta^{4}(r)}+
A^{2}\frac{1}{L^{2}}\frac{1}{\Delta^{2}(r)}}}\nonumber\\
& = &
\frac{L}{T}\frac{\pi}{2}\int_{0}^{r}\left|
\frac{d}{dr'}\left(\frac{Z(r')\Delta(r')}{B\frac{\Gamma^{2}(r')}{\Delta(r')}
\pm iA}\right)\right| dr'\nonumber\\ & &+\frac{L}{T}\int_{0}^{r}
\frac{Z^{2}(r')\Delta(r')}{\sqrt{B^{2}\frac{\Gamma^{4}(r')}{\Delta^{2}(r')}+
A^{2}}}dr'\nonumber\\
& & +\frac{L}{T}\frac{\pi}{2}
\frac{Z(r)\Delta(r)}{\sqrt{B^{2}\frac{\Gamma^{4}(r)}{\Delta^{2}(r)}+
A^{2}}}.
\end{eqnarray}
By using Eq.~(\ref{zetadef}) it follows that
\begin{eqnarray}
|\widetilde{I}_{\pm}(r)| & \leq &
\frac{L}{T}\frac{\pi}{2}\int_{0}^{r}\left|\frac{d}{dr'}
\left(\frac{Z(r')\Delta(r')}{B\frac{\Gamma^{2}(r')}{\Delta(r')}\pm
iA}\right)\right| dr'\nonumber \\ 
& & +\frac{L}{T}\frac{1}{\sqrt{B^{2}\zeta +
A^{2}}}\int_{0}^{r}Z^{2}(r')\Delta(r')dr'\nonumber\\
& & +\frac{L}{T}\frac{\pi}{2}\frac{Z(r)\Delta(r)}{\sqrt{B^{2}\zeta +
A^{2}}}.
\end{eqnarray}
One may use $Z(r)\Delta(r) \leq \sqrt{(N-1)/N}$ and
 Eq.~(\ref{Zegensk}) to obtain
\begin{eqnarray}
\label{tiIabs}
|\widetilde{I}_{\pm}(r)| & \leq & 
\frac{L}{T}\frac{\pi}{2}\int_{0}^{r}
\left|\frac{d}{dr'}\left(\frac{Z(r')\Delta(r')}{B\frac{\Gamma^{2}(r')}{\Delta(r')}\pm iA}\right)\right| dr'\nonumber\\
 & &+\frac{\pi
L}{T}\sqrt{\frac{N-1}{N}}\frac{1}{\sqrt{B^{2}\zeta^{2}+ A^{2}}}.
\end{eqnarray}
The derivative contained in the integrand on the right-hand side of the
above equation is performed by first multiplying both the numerator
and the denominator by $\Delta(r')$ and then using
$Z(r)\Delta^{2}(r) =\sqrt{N-1}/N$. Moreover, we use
Eq.~(\ref{zetadef}). The result is
\begin{eqnarray}
\label{absderiv}
 & &\int_{0}^{r}\left|
\frac{d}{dr'}\left(\frac{Z(r')\Delta(r')}{B\frac{\Gamma^{2}(r')}{\Delta(r')}\pm
iA}\right)\right| dr'\nonumber\\ &\leq &\frac{ B}{B^{2}\zeta^{2} +
A^{2}} \int_{0}^{1}Z(r')\left|\frac{d}{dr'}\Gamma^{2}(r')\right| dr'
\nonumber\\ & & + \frac{A}{B^{2}\zeta^{2} +
A^{2}}\int_{0}^{1}Z(r')\left|\frac{d}{dr'}\Delta(r')\right|dr'.
\end{eqnarray}
Note that in the inequality above we have extended the integration
interval from $[0,r]$ to $[0,1]$. By combining Eqs.~(\ref{condition}),
(\ref{ZderD}), (\ref{Ldef}), (\ref{tiIabs}), and (\ref{absderiv}) one
obtains
\begin{eqnarray}
\label{wtI}
|\widetilde{I}(r)| & \leq 
&\frac{\pi^{2}}{2}\frac{\sqrt{N}}{T}\Bigg(\frac{1}{\sqrt{B^{2}\zeta^{2}+
A^{2}}} + \frac{A}{B^{2}\zeta^{2} + A^{2}}\Bigg)\nonumber\\ &
&+\frac{\pi^{2}}{4}
\frac{\sqrt{N}}{T}\sqrt{\frac{N}{N-1}}\frac{BK}{B^{2}\zeta^{2} + A^{2}}.
\end{eqnarray}

\subsubsection{The term $\widetilde{J}$}
As in the case of global search one obtains $|\widetilde{J}(r)| \leq
|\widetilde{\rho}_{10}(0)|\,\,|\widetilde{G}_{+}[1](r)|$, with
$\widetilde{G}$ as in Eq.~(\ref{Gdef}) with $g(r) \equiv 1$, and $Q$
and $R$ replaced by $\widetilde{Q}$ and $\widetilde{R}$. Note that
$\widetilde{\rho}(0)=\rho(f^{-1}(0))= \rho(0)$ and thus
$\widetilde{\rho}_{10}(0) = \rho_{10}(0)$.  Using similar arguments as
in the previous sections, one obtains
\begin{eqnarray}
|\widetilde{G}_{+}[1](r)| &  \leq &
\frac{L}{T}\int_{0}^{r}
\left|\frac{d}{dr'}\left(\frac{Z(r')\Delta(r')}{B\frac{\Gamma^{2}(r')}{\Delta(r')}+ iA}\right)\right|dr'\nonumber\\ 
& &+\frac{L}{T}\frac{\sqrt{N-1}}{N}\frac{1}{\sqrt{B^{2}\Gamma^{4}(0)+
A^{2}}}\nonumber\\ 
& &+\frac{L}{T}\frac{Z(r)\Delta(r)}{\sqrt{B^{2}\zeta^{2}+
A^{2}}}.
\end{eqnarray}
Next we use $Z(r)\Delta(r) \leq \sqrt{(N-1)/N}$ together with
Eqs.~(\ref{condition}), (\ref{ZderD}), (\ref{Ldef}), and
(\ref{absderiv}), yielding
\begin{eqnarray}
\label{wtJ}
|\widetilde{J}(r)| &\leq &\frac{\pi}{2}
|\widetilde{\rho}_{10}(0)|\frac{\sqrt{N}}{T}
\frac{1}{\sqrt{B^{2}\zeta^{2}+ A^{2}}}\nonumber\\ & & +
\frac{\pi}{2}|\widetilde{\rho}_{10}(0)|
\frac{1}{T}\frac{1}{\sqrt{B^{2}\Gamma^{4}(0)+
A^{2}}}\nonumber\\ & & +\frac{\pi}{2}
|\widetilde{\rho}_{10}(0)|\frac{\sqrt{N}}{T}\frac{BK}{B^{2}\zeta^{2} +
A^{2}}\sqrt{\frac{N}{N-1}}\nonumber\\ & & +\frac{\pi}{2}
|\widetilde{\rho}_{10}(0)|\frac{\sqrt{N}}{T}\frac{2A}{B^{2}\zeta^{2} +
A^{2}}.
\end{eqnarray}

\subsubsection{Collecting the results}
We combine Eqs.~(\ref{wtI}) and (\ref{wtJ}) together with
$|\widetilde{Y}(0)-\widetilde{Y}(r)|\leq 4|\widetilde{I}(r)| +
4|\widetilde{J}(r)|$, $A>0$, $B^{2}\zeta^{2} + A^{2} \geq A^{2}$, and
$B^{2}\Gamma^{4}(0)+ A^{2}\geq A^{2}$. Furthermore, we assume $N\geq
2$, which gives $\sqrt{N/(N-1)}\leq\sqrt{2}$, and results in \cite{error}
\begin{eqnarray}
\label{soloc}
|\widetilde{\rho}_{00}(0)-\widetilde{\rho}_{00}(r)| & \leq &
\left(|\widetilde{\rho}_{10}(0)|+\frac{\pi}{2}\right)\frac{\sqrt{2}\pi
BK}{A^{2}}\frac{\sqrt{N}}{T}\nonumber\\ 
 & & +(3|\widetilde{\rho}_{10}(0)| + 2\pi)
\frac{\pi}{A}\frac{\sqrt{N}}{T}\nonumber\\
& & +|\widetilde{\rho}_{10}(0)|\frac{\pi}{A}\frac{1}{T}.
\end{eqnarray}
The leading term of this equation is proportional to
$\sqrt{N}/T$. Hence, a sufficient condition for adiabaticity is $T\gg
\sqrt{N}$.  Equation (\ref{soloc}) can be expressed in terms of the
solution $\rho(s)$ of Eq.~(\ref{startlocal}) by the substitution
$r=f(s)$. However, this change of variables does not affect the
condition for adiabaticity.

\subsection{\label{sec:condition}The condition}
Both for the global and local search we have restricted the
decoherence term of the master equation to fulfill the condition in
Eq.~(\ref{condition}). This condition puts a limit on the fluctuations
of $\Gamma$ with respect to $s$, as a function of the list-length
$N$. However, one may note that the condition can be relaxed, in the
sense that we may allow $K$ to be dependent on $N$. By inserting
$K=cN^{a}$, where $c\geq 0$ and $a\geq 0$ are constants, into
Eqs.~(\ref{sogl}) and (\ref{soloc}), one finds that the new scaling
becomes $N^{1+a}$ and $N^{1/2+a}$ for the global and local search,
respectively. Thus, the scalings get worse, but one finds that there
is a ``window'' $0 < a <1/2$ where the local search still outperforms the
classical search, although not with the optimal efficiency of the
Grover search \cite{Grover}.

In the following we attempt to obtain an indication of to what extent the
condition in Eq.~(\ref{condition}) is restrictive or not. For the sake
of simplicity we do so for the more restrictive case where $K$ is
independent of $N$. We assume $\Gamma(s) =
\eta\bm{(}\Delta(s)\bm{)}$, where $\eta:(0,\infty) \rightarrow
(0,\infty)$, is an increasing and sufficiently smooth function. Now, a
larger value of $B\Gamma^{2}(s)$ implies stronger decoherence. Thus,
we have chosen to consider a decoherence model for which the strength
of the decoherence increases with the energy difference between the
two instantaneous eigenstates, but we do not specify any details on
how it grows. We intend to derive a sufficient condition on
$\eta$ for $\Gamma$ to fulfill the condition in Eq.~(\ref{condition}).

By the assumptions on $\eta$ and the form of $\Delta$, it follows that
$\Gamma$ is decreasing on the interval $[0,1/2]$ and increasing on
$[1/2,1]$. Partial integration implies
\begin{eqnarray}
\int_{0}^{1}Z(s)\left|\frac{d}{ds}\Gamma^{2}(s)\right|ds  & =
& 2\int_{1/2}^{1}Z(s)\frac{d}{ds}\eta^{2}\bm{(}\Delta(s)\bm{)}
ds\nonumber\\ 
& = & -2\sqrt{N-1}\eta^{2}(N^{-1/2})\nonumber\\
& & -2\int_{1/2}^{1}\eta^{2}\bm{(}\Delta(s)\bm{)}\frac{d}{ds}Z(s)ds\nonumber\\
& & + 2\frac{\sqrt{N-1}}{N}\eta^{2}(1). 
\end{eqnarray}
We use $\eta^{2}(N^{-1/2})\geq 0$ and
$2\eta^{2}(1)\sqrt{N-1}/N\leq 2\eta^{2}(1)$. Moreover, we use
$Z(s)\Delta^{2}(s) = \sqrt{N-1}/N$, which leads to
\begin{eqnarray}
\int_{0}^{1}Z(s)\left|\frac{d}{ds}\Gamma^{2}(s)\right|ds  & \leq & 
4\frac{\sqrt{N-1}}{N}\int_{1/2}^{1}
\frac{\eta^{2}\bm{(}\Delta(s)\bm{)}}{\Delta^{3}(s)}\frac{d\Delta}{ds}ds
\nonumber \\ & & + 2\eta^{2}(1).
\end{eqnarray}
Only the integral on the right-hand side of the above
expression may potentially grow when $N$ increases. If we treat
this term separately we find
\begin{equation}
\frac{\sqrt{N-1}}{N}\int_{1/2}^{1}
\frac{\eta^{2}\bm{(}\Delta(s)\bm{)}}{\Delta^{3}(s)}\frac{d\Delta}{ds}ds
\leq
\frac{1}{\sqrt{N}}\int_{1/\sqrt{N}}^{1}
\frac{\eta^{2}(\Delta)}{\Delta^{3}}d\Delta.\nonumber
\end{equation}
This implies that a sufficient condition for the inequality in
Eq.~(\ref{condition}) to hold, is that the expression on the
right-hand side of the above equation, is bounded. From the
assumptions on $\eta$ it follows that the limiting behavior of this
integral is determined by the asymptotic behavior of the function
$\eta$ as $\Delta \rightarrow 0$.  Assume that $\eta(x)\leq
Cx^{\sigma}$ for some constants $C$ and $\sigma\geq 0$. Then it
follows that
\begin{equation}
\label{uppd}
\frac{1}{\sqrt{N}}\int_{\frac{1}{\sqrt{N}}}^{1}\frac{\eta^{2}(x)}{x^{3}}dx 
\leq  \left\{ \begin{array}{ll}
\frac{C^{2}}{2}\frac{\ln N}{\sqrt{N}}, & \sigma = 1,\\
\frac{C^{2}}{2(\sigma-1)}\left(N^{-\frac{1}{2}}-N^{\frac{1}{2}-\sigma}\right),
 & \sigma \neq 1.
\end{array}\right.
\end{equation}
In the case where $\sigma = 1$, the function $\Gamma$ satisfies the
inequality in Eq.~(\ref{condition}), since $(\ln N)/\sqrt{N}$ is a
bounded function on the nonzero natural numbers.  In the case where
$\sigma \neq 1$, one can see that the right-hand side of
Eq.~(\ref{uppd}) contains two terms. None of these terms increase in
magnitude, with increasing $N$, if $\sigma\geq 1/2$. Hence, we may
conclude that for all $\sigma \geq 1/2$ the condition in
Eq.~(\ref{condition}) is fulfilled. In other words, for a sufficiently
smooth, positive, and increasing function $\eta$, the only condition
we have required is $\eta(x)\leq Cx^{\sigma}$ with $\sigma \geq 1/2$.

\section{\label{sec:wopen}The wide-open case with
 $\Gamma(s) = \Delta^{\sigma}(s)$}
The wide-open case is obtained if we let $A = 0$ (and $B=1$ for
convenience) which in the global search case yields the master
equation
\begin{equation}
\label{widoppen}
\frac{d}{ds}\rho(s) = -T\bm{[}W(s),[W(s),\rho(s)]\bm{]}
\end{equation}
and in the local search case yields 
\begin{equation}
\label{widoppenlocal}
\frac{d}{dr}\widetilde{\rho}(r) 
= -T\frac{df^{-1}}{dr}\bm{[}W(r),[W(r),\widetilde{\rho}(r)]\bm{]},
\end{equation}
where we have changed variables to $r=f(s)$ as described in
Sec.~\ref{sec:semiopenlocal}.
 
Even with the assumption that $W(s)$ should be simultaneously
diagonalizable with $H(s)$, we still have a rather large freedom in
the choice of $W$, in terms of how $\Gamma$ depends on the parameter
$s$ and the list-length $N$.  In the previous sections it was only
necessary to require that $\Gamma$ should fulfill the condition
in Eq.~(\ref{condition}). In this case, however, we need to specify more
about $W$.

Our main object of study in this section is the case $W(s) = H(s)$,
which corresponds to $\Gamma(s) = \Delta(s)$. However, in order to
prove that in the wide-open case the asymptotic behavior is dependent
on the choice of $W$, we consider the generalization $\Gamma(s) =
\Delta^{\sigma}(s)$.
Note in particular that if $\sigma = n$, where $n$ is an odd natural
number, then $\Gamma(s) = \Delta^{n}(s)$ can be obtained with $W(s) =
2^{n-1}(H(s) + (1/2)\hat{1})^{n}$.

Here we prove that in the wide-open case, with
$\Gamma(s)=\Delta^{\sigma}(s)$, the global and local conditions for
adiabaticity are $T\gg N^{\sigma+1/2}$ and $T\gg N^{\sigma}$,
respectively. We first prove that these are sufficient
conditions. Thereafter follows a proof of necessity.

\subsection{\label{sec:globsuff}Global search: Sufficiency}
Let $A = 0$ and $B=1$ in Eq.~(\ref{tvaanivaa}) and assume that
$\Gamma(s) = \Delta^{\sigma}(s)$, which gives
\begin{equation}
Q(s) = \int^{s}_{0}\Delta^{2\sigma}(s')ds'.
\end{equation}
Assume $\sigma \geq 0$, and use $\Delta(s)\geq \delta= 1/\sqrt{N}$.
It follows that $Q(s)\geq N^{-\sigma}s$, for all $s\geq 0$.  Moreover, 
 since $s'\geq s''$, it follows that $Q(s')-Q(s'')\geq
N^{-\sigma}(s'-s'')$. This can be used, together with $|Y(s)|\leq 1$,
to show that Eq.~(\ref{tvaanivaa}) leads to
\begin{eqnarray}
\label{sdghn}
|Y(0)-Y(s)|& \leq &
4\int_{0}^{s}\!\!\!\int_{0}^{s'}
\!\!\!e^{-TN^{-\sigma}(s'-s'')}Z(s')Z(s'')ds''ds'\nonumber\\
& &+ 4|\rho_{10}(0)|\int_{0}^{s}e^{-TN^{-\sigma}s'}Z(s')ds'. 
\end{eqnarray}
One may use $Z(s)\leq \sqrt{N-1}$, integrate, and use
Eq.~(\ref{Zegensk}) to obtain
\begin{equation}
\label{globwo}
 |\rho_{00}(0)-\rho_{00}(s)| 
\leq   (2|\rho_{10}(0)|+\pi)\frac{N^{\sigma+1/2}}{T}.
\end{equation}
Thus, we have proved that a sufficient condition for $\rho_{00}(s)$ to
approach $\rho_{00}(0)$ is $T \gg N^{\sigma +1/2}$, if $\sigma\geq 0$.

\subsection{\label{sec:locsuff}Local search: Sufficiency}
Here we require $\sigma$ to fulfill $\sigma \geq 1$.  As before we let
$A = 0$ and $B=1$ in Eq.~(\ref{tvaanivaa}), but we replace $Q$ with
$\widetilde{Q}$ defined as
\begin{eqnarray}
\label{wQWH}
\widetilde{Q}(r) & = &
\int_{0}^{r}\Delta^{2\sigma}(r')\frac{df^{-1}}{dr'}dr'\nonumber\\
& = & \frac{1}{L}\int_{0}^{r}\Delta^{2\sigma-2}(r')dr'\nonumber\\ 
& \geq & \frac{1}{LN^{\sigma-1}}r,
\end{eqnarray}
with $L$ as in Eq.~(\ref{Ldef}), and where we in the inequality have
used $\Delta(s)\geq \delta = 1/\sqrt{N}$ and the assumption $\sigma \geq 1$.

Inserting Eq.~(\ref{wQWH}) into Eq.~(\ref{tvaanivaa}) yields 
\begin{eqnarray}
\label{olikhet}
& &|\widetilde{Y}(0)-\widetilde{Y}(r)|\nonumber\\ & \leq &
4\int_{0}^{r}\int_{0}^{r'}
e^{-(r'-r'')T/(LN^{\sigma-1})}Z(r')Z(r'')dr''dr'\nonumber\\
& &+4|\widetilde{\rho}_{10}(0)|\int_{0}^{r}e^{-r'T/(LN^{\sigma-1})}Z(r')dr'.
\end{eqnarray}
By a reasoning very similar to the derivation leading from
Eq.~(\ref{sdghn}) to Eq.~(\ref{globwo}) it follows that
\begin{eqnarray}
|\widetilde{Y}(0)-\widetilde{Y}(r)| & \leq &
2\pi\frac{LN^{\sigma-1}\sqrt{N-1}}{T}\nonumber\\
& & +
4|\widetilde{\rho}_{10}(0)|\frac{LN^{\sigma-1}\sqrt{N-1}}{T}, 
\end{eqnarray}
which, together with Eq.~(\ref{Ldef}), gives
\begin{equation}
\label{wioloc}
|\widetilde{\rho}_{00}(0)-\widetilde{\rho}_{00}(r)| 
\leq  (2|\widetilde{\rho}_{10}(0)| + \pi)\frac{\pi}{2}\frac{N^{\sigma}}{T}.
\end{equation}
Thus, a sufficient condition for adiabaticity is $T \gg N^{\sigma}$.
 
\subsection{\label{sec:necess}Necessity}
In Secs.~\ref{sec:globsuff} and \ref{sec:locsuff} we proved
sufficient conditions for conservation of the probability to find
the system in the ground state, throughout the evolution. Here
we prove that these sufficient conditions are also necessary, but
under the simplifying assumption that we start in the ground state of
the system. Moreover, we focus
on the end point $s=1$, i.e., we consider the success
probability $p=\rho_{00}(1)$, and necessary conditions for $p$ to
approach unity.

Due to the effectively two-dimensional Hilbert space of the problem we
may use a Bloch vector representation in terms of which the state of
the system and the Hermitian operator $W$ read
\begin{eqnarray}
\rho(s) &=& \frac{1}{2}[\hat{1}+\bm{v}(s)\cdot\bm{\sigma}],\\
W(s) &= &\frac{1}{2}\{[w_{0}(s)+w_{1}(s)]\hat{1}
+\bm{u}(s)\cdot{\bm\sigma}\},
\end{eqnarray}
where $\bm{v}$ and $\bm{u}$ are elements in $\mathbb{R}^{3}$,
$||\bm{v}||\leq 1$, and $\bm{\sigma} =
(\sigma_{x},\sigma_{y},\sigma_{z})$ are the Pauli spin operators.
It follows that $\bm{u}(s) =
-\Gamma(s)\bm{q}(s)$, where $\bm{q}(s)$ is the Bloch unit-vector
representing the ground state of
$H(s)$.

If we use the assumption $\Gamma(s) = \Delta^{\sigma}(s)$, then
Eq.~(\ref{widoppen}) can be rewritten in the Bloch representation
 as
\begin{equation}
\label{blochekv}
\dot{\bm{v}} = 
-T\xi(s)\{\bm{v}-\bm{q}(s)[\bm{q}(s)\cdot{\bm{v}}]\},
\end{equation}
where
\begin{equation}
\xi(s) = \Delta^{2\sigma}(s),\quad \sigma \geq 0.
\end{equation}
In the local case Eq.~(\ref{widoppenlocal}) again becomes
Eq.~(\ref{blochekv}), but with the parameter $s$ replaced by $r$ and
\begin{equation}
\xi(r) = \Delta^{2\sigma}(r)\frac{df^{-1}}{dr}
 = \frac{\Delta^{2\sigma-2}(r)}{L},\quad \sigma\geq 1.
\end{equation}
Most of the material presented in this subsection is independent of
whether we consider the global or the local search. The difference
is exclusively carried by the function $\xi$. Due to this we will in
the rest of this subsection use the variable $s$, although we should,
to be correct, use $r$ in the local search case. In
Sec.~\ref{sec:locnecess} we properly use the variable $r$ again.

Let $|\psi\rangle$ and $|\overline{\psi}\rangle$ defined in
Eq.~(\ref{basis}) correspond to the north and south pole of the Bloch
sphere, respectively. The Bloch vector $\bm{q}$, corresponding to the
instantaneous ground state, can then be expressed as
\begin{equation}
\label{qutr}
\bm{q}(s) = \frac{1}{\Delta(s)}
\left(2s\frac{\sqrt{N-1}}{N}, 0,1-2s\frac{N-1}{N}\right).
\end{equation}
Note that $\bm{q}$ moves in the $x$-$z$ plane only, and that $\bm{q}(0) =
(0,0,1)$. Let $\theta(s)$ be the angle between $\bm{q}(s)$ and
$\bm{q}(0)$. This angle can be expressed as
\begin{equation}
\label{Ynoll}
\cos\bm{(}\theta(s)\bm{)} = \bm{q}(s)\cdot\bm{q}(0) =
 \frac{1-(N-1)(2s-1)}{\sqrt{N}\sqrt{1 + (N-1)(2s-1)^{2}}}.
\end{equation}
For $N\geq 2$ the right-hand side of the above equation is strictly
decreasing on the interval $0\leq s\leq 1$, beginning at the value $1$
and ending at the value $2/N-1$.  It follows that we can choose
$\theta(s)$ as an increasing function with range in $[0,\pi]$.

Since the system is started in the ground state $\rho(0) =
|E_{0}(0)\rangle\langle E_{0}(0)|$, the initial condition is
$\bm{v}(0) =\bm{q}(0)$. Since both $\bm{q}(s)$ and $\bm{v}(0)$ are
fixed to the  $x$-$z$ plane it follows, from the form of
Eq.~(\ref{blochekv}), that the motion of $\bm{v}$ is fixed to the $x$-$z$ 
plane. Thus, we may restrict our analysis to this plane only.

\begin{Lemma}
\label{lemins}
Let $\bm{q}_{0}$, $\bm{q}'$, and $\bm{q}''$ be
three unit vectors in $\mathbb{R}^{2}$, and let $\bm{v}$ be such
that $||\bm{v}||\leq 1$. Suppose that the angle between
$\bm{q}_{0}$ and $\bm{q}''$ is greater than or equal to the
angle between $\bm{q}_{0}$ and $\bm{q}'$. Moreover, let both
these angles be in the interval $[0,\pi]$. Then
\begin{equation}
\label{inputas}
\bm{v}\cdot\bm{q}'\geq \bm{q}_{0}\cdot\bm{q}'
\end{equation}
implies
\begin{equation} 
\label{outputas}
\bm{v}\cdot\bm{q}''\geq \bm{q}_{0}\cdot\bm{q}''.
\end{equation} 
\end{Lemma}

\textit{Proof.}
Consider a coordinate system such that $\bm{q}_{0}$ corresponds to
$(1,0)$, $\bm{q}'$ to $(\cos\theta_{1},\sin\theta_{1})$, and
$\bm{q}''$ to $(\cos\theta_{2},\sin\theta_{2})$, with $0\leq
\theta_{1}\leq
\theta_{2}\leq \pi$. In the cases when
$\theta_{1},\theta_{2}$ assumes possible combinations of $0,\pi$,
the lemma is trivially true. Hence, we assume $0<\theta_{1}\leq
\theta_{2}<\pi$.

If we let $\bm{v} = (x,y)$, then Eq.~(\ref{inputas}) becomes
$x\cos\theta_{1} + y\sin\theta_{1}\geq \cos\theta_{1}$. By using
 $\sin\theta_{1}>0$, this can be rewritten as
\begin{equation}
\label{sfgn}
y\geq (1-x)\cot\theta_{1}.
\end{equation} 
Since cotangent is a  decreasing function on the interval $(0,\pi)$, and
since $1-x\geq 0$, it follows that the right-hand side of
Eq.~(\ref{sfgn}) is larger than or equal to
$(1-x)\cot\theta_{2}$. The above procedure can be
reversed to obtain Eq.~(\ref{outputas}). $\Box$\newline

\begin{Lemma}
\label{lemmabloch}
If $\bm{v}(s)$ is the solution of Eq.~(\ref{blochekv}) with
initial condition $\bm{v}(0)= \bm{q}(0)$, and with
$\bm{q}(s)$ as in Eq.~(\ref{qutr}), then
\begin{equation}
\label{basolikhet}
\bm{v}(s)\cdot\bm{q}(s)\geq \bm{q}(0)\cdot\bm{q}(s),\quad \forall s\in[0,1].
\end{equation}
\end{Lemma}

\textit{Proof.}
The strategy of this proof is to make a Cauchy-Euler polygon
approximation of Eq.~(\ref{blochekv}), and show an inequality similar
to Eq.~(\ref{basolikhet}) for every step size. Since the solution
$\bm{v}(s)$ is obtained in the limit of small step sizes, the solution
will satisfy the inequality in Eq.~(\ref{basolikhet}).

Let $(s_{k})_{k=0}^{M}$ be a partition of the interval $[0,1]$ such
that $s_{0} = 0$. The partition has the step size $D\!s = 1/M$.  Let
$\bm{q}_{k} = \bm{q}(s_{k})$ and let
\begin{eqnarray}
\label{xgn}
\bm{v}_{0} & =& \bm{q}_{0},\nonumber\\
\bm{v}_{k+1} & = &\bm{v}_{k} - 
D\!s T\xi(s_{k})[\bm{v}_{k}-\bm{q}_{k}(\bm{q}_{k}\cdot\bm{v}_{k})  ].
\end{eqnarray} 
Let $\bm{v}^{(D\!s)}(s)$ be the linear interpolation of
$\bm{v}_{0},\bm{v}_{1},\ldots,\bm{v}_{k}$, meaning that on the
interval $[s_{k},s_{k+1}]$ the function $\bm{v}^{(D\!s)}(s)$ is
defined by
\begin{equation}
\bm{v}^{(D\!s)}(s) =
\bm{v}_{k+1}\frac{s-s_{k}}{s_{k+1}-s_{k}} +
\bm{v}_{k}\frac{s_{k+1}-s}{s_{k+1}-s_{k}}.
\end{equation}

Now we prove that the approximation $\bm{v}^{(D\!s)}(s)$ converges
uniformly on the interval $[0,1]$ to the solution $\bm{v}(s)$, as
$D\!s\rightarrow 0$.
Define 
\begin{equation}
\bm{f}(s,\bm{v}) =
-T\xi(s)\{\bm{v}-\bm{q}(s)[\bm{q}(s)\cdot\bm{v}]\}.
\end{equation}
In both the global and the local case the function $\bm{f}$ can be
shown to be continuous on $\mathbb{R}\times\mathbb{R}^{3}$, and
to have a continuous partial derivative with respect to $s$.
If one uses the assumption $\sigma\geq 0$ in the global case, and
$\sigma\geq 1$ as well as the additional assumption $N\geq 2$ in the
local case, one finds
\begin{equation}
\label{xicond}
0\leq \xi(s)\leq 1.
\end{equation} 
It follows that $\bm{f}$ fulfills the
 Lipschitz condition
\begin{equation}
\label{Lipschitz}
||\bm{f}(s,\bm{x})-\bm{f}(s,\bm{y})|| \leq 
 T||\bm{x}-\bm{y}||,\quad (s,\bm{x}),(s,\bm{y})\in[0,1]\times\mathbb{R}^{3}.
\end{equation}
If we assume that $D\!s \leq 1/T$, then 
Eq.~(\ref{xicond}) implies
\begin{equation}
\label{instangn}
|1-D\!sT\xi(s)|\leq 1,\quad \forall s\in [0,1].
\end{equation}
By induction using Eqs.~(\ref{xgn}) and (\ref{instangn}) one can show
that $||\bm{v}_{k}||\leq 1$.  This in turn implies that the linear
interpolation $\bm{v}^{(D\!s)}(s)$ fulfills
\begin{equation}
\label{spifub}
||\bm{v}^{(D\!s)}(s)||\leq 1,\quad \forall s\in[0,1].
\end{equation}
The inequality $||\bm{v}_{k}||\leq 1$, together with Eq.~(\ref{xgn}),
can be used to show that
\begin{equation}
||\bm{v}_{k+1}-\bm{v}_{k}||\leq D\!s T\xi(s_{k})||\bm{v}_{k}||\leq
(s_{k+1}-s_{k})T.
\end{equation}
This gives
\begin{equation}
\label{mgkhn}
||\bm{v}^{(D\!s)}(s'')-\bm{v}^{(D\!s)}(s')|| \leq T|s''-s'|,\quad
s',s''\in [0,1].
\end{equation}

The fact that the partial derivative of $\bm{f}$, with respect to $s$, is
continuous on $\mathbb{R}\times\mathbb{R}^{3}$, implies that
there exists a constant $C$ such that
\begin{equation}
\label{mdfs}
\begin{split}
||\bm{f}(s'',\bm{v})-\bm{f}(s',\bm{v})||\leq C|s''-s'|,
\end{split}
\end{equation}
for all $s',s''\in [0,1]$, and all
$\bm{v}\in\mathbb{R}^{3}:||\bm{v}||\leq 1$.  By combining
Eqs.~(\ref{Lipschitz})-(\ref{mdfs}) one obtains
\begin{equation}
||\bm{f}(s'',\bm{v}^{(D\!s)}(s''))-\bm{f}(s',\bm{v}^{(D\!s)}(s'))||
 \leq (T^{2}+C)|s''-s'|,
\end{equation}
for all $s',s''\in[0,1]$.
This can be used to show that 
\begin{equation}
\label{eapproximativ}
 \left|\left|\bm{v}^{(D\!s)}(s'')-\bm{v}^{(D\!s)}(s')- \int_{s'}^{s''}
\!\!\!\!\bm{f}(s,\bm{v}^{(D\!s)}(s)) ds  \right|\right| \leq \epsilon|s''-s'|,
\end{equation}
for all $s',s''\in [0,1]$, and where $\epsilon = D\!s(T^{2}+C)$.
Since both $\bm{f}$ and $\bm{v}^{(D\!s)}$ are continuous,
Eq.~(\ref{eapproximativ}) implies that $\bm{v}^{(D\!s)}$ is an
$\epsilon$-approximative solution \cite{BiRo, Amann} to
Eq.~(\ref{blochekv}).

Since $\bm{f}$ is continuous and fulfills the Lipschitz condition
in Eq.~(\ref{Lipschitz}) it follows \cite{BiRo, Amann} from
Eq.~(\ref{eapproximativ}) that
\begin{eqnarray}
\label{nldsvk}
||\bm{v}^{(D\!s)}(s)-\bm{v}(s)|| 
&\leq & ||\bm{v}^{(D\!s)}(0)-\bm{v}(0)||e^{Ts}\nonumber\\
& &  + D\!s\frac{(T^{2}+C)}{T}(e^{Ts}-1),\nonumber\\
\end{eqnarray}
for all $s\in [0,1]$.  Since $\bm{v}(0) = \bm{v}^{(D\!s)}(0) =
\bm{q}(0)$ it follows from Eq.~(\ref{nldsvk}) that
$\bm{v}^{(D\!s)}(s)$ converges uniformly on $[0,1]$ to the solution
$\bm{v}(s)$, as $D\!s\rightarrow 0$. (One may note that the solution
$\bm{v}(s)$ exists and is unique, since $\bm{f}$ is continuous and
Lipschitz on $[0,1]\times\mathbb{R}^{3}$.) The solution $\bm{v}(s)$ is
continuous as it is the limit function of the continuous and uniformly
converging functions $\bm{v}^{(D\!s)}(s)$ on the subset $[0,1]$ of
$\mathbb{R}$. (See Theorem 7.12 in Ref.~\cite{Rudin}.)

In the following we focus on the proof of the inequality
in Eq.~(\ref{basolikhet}).  We use the fact that the vectors $\bm{v}_{k}$
and $\bm{q}_{k}$ are restricted to the  $x$-$z$ plane of the Bloch
sphere. By induction we prove that $\bm{v}_{k}\cdot\bm{q}_{k}\geq
\bm{q}_{0}\cdot\bm{q}_{k}$.

By using Eq.~(\ref{xgn}) one finds
\begin{equation}
\bm{v}_{k+1}\cdot\bm{q}_{k} = \bm{v}_{k}\cdot\bm{q}_{k}
.
\end{equation}
If this is combined with the induction hypothesis
$\bm{v}_{k}\cdot\bm{q}_{k}\geq
\bm{q}_{0}\cdot\bm{q}_{k}$
we obtain
\begin{equation}
\label{enbn}
\bm{v}_{k+1}\cdot\bm{q}_{k} \geq \bm{q}_{0}\cdot\bm{q}_{k}.
\end{equation}
The angle between $\bm{q}_{k+1}$ and $\bm{q}_{0}$ is larger than the
angle between $\bm{q}_{k}$ and $\bm{q}_{0}$, and these angles are in
the range $[0,\pi]$. According to Lemma \ref{lemins} with $\bm{q}'
=\bm{q}_{k}$ and $\bm{q}'' = \bm{q}_{k+1}$, it follows from
Eq.~(\ref{enbn}) that
\begin{equation}
\bm{v}_{k+1}\cdot\bm{q}_{k+1} \geq \bm{q}_{0}\cdot\bm{q}_{k+1}.
\end{equation}
Since the induction hypothesis is true for $k = 0$, we obtain
\begin{equation}
\label{olik}
\bm{v}_{k}\cdot\bm{q}_{k}\geq \bm{q}_{0}\cdot\bm{q}_{k},
\end{equation}
for all $0\leq k\leq M$.
This is true regardless of the choice of step size $D\!s$.

Suppose there exists a point $\widetilde{s}$ for which the inequality
$\bm{v}(s)\cdot\bm{q}(s)\geq
\bm{q}(0)\cdot\bm{q}(s)$ does not hold. Then there exists a
number $a>0$, such that
$[\bm{v}(\widetilde{s})-\bm{q}(0)]\cdot
\bm{q}(\widetilde{s}) = -a <0$. By continuity of $\bm{v}$
and $\bm{q}$, there must exist an interval
$\widetilde{s}_{1}<\widetilde{s}<\widetilde{s}_{2}$ such that
\begin{equation}
\label{ineq}
[\bm{v}(s)-\bm{q}(0)]\cdot \bm{q}(s)\leq -a/2,\quad
\forall s\in[\widetilde{s}_{1},\widetilde{s}_{2}].
\end{equation}
Now consider approximations $\bm{v}^{(D\!s)}(s)$ in this
interval.
\begin{eqnarray}
[\bm{v}^{(D\!s)}(s)-\bm{q}(0)]\cdot \bm{q}(s) &
\leq &  [\bm{v}^{(D\!s)}(s)-\bm{v}(s)]\cdot
\bm{q}(s) -a/2 \nonumber\\ & \leq & \sup_{s\in
[\widetilde{s}_{1},\widetilde{s}_{2}]}\big|\big|\bm{v}^{(D\!s)}(s)
-\bm{v}(s)\big|\big|\nonumber\\
 & & -a/2.
\end{eqnarray}
Since $\bm{v}^{(D\!s)}(s)$ converges uniformly to $\bm{v}(s)$, as
$D\!s\rightarrow 0$, there exists a sufficiently small $D\!s$ such
that $\sup||\bm{v}^{(D\!s)}(s)-\bm{v}(s)|| < a/2$. Thus, for a
sufficiently small $D\!s$ it follows that
$[\bm{v}^{(D\!s)}(s)-\bm{q}(0)]\cdot
\bm{q}(s)<0$, for all $s\in
[\widetilde{s}_{1},\widetilde{s}_{2}]$. This contradicts Eq.~(\ref{olik}),
since we may choose $s$ to be some element of the partition
$(s_{k})_{k=0}^{M}$ inside the interval
$[\widetilde{s}_{1},\widetilde{s}_{2}]$. (Such an element exists if
$D\!s$ is sufficiently small.) Hence, by contradiction we have proved
the statement of the lemma.  $\Box$\newline

\subsubsection{\label{sec:globnecess}Global search}
In the following we let $Y_{T}$ denote the solution of
Eq.~(\ref{tvaanivaa}) for a given run time $T$, with $A=0$ and $B=1$.
Similarly we let $\rho_{T}$ denote the solution of
Eq.~(\ref{widoppen}), for the run time $T$.

\begin{Lemma}
\label{lemmatre}
\begin{equation}
\label{Yineq}
Y_{T}(s)\geq Y_{0}(s),\quad T\geq 0,\quad 0\leq s\leq 1.
\end{equation}
\end{Lemma}

\textit{Proof.}
If  $\rho_{T}(0)=|E_{0}(0)\rangle\langle E_{0}(0)|$,
then, by construction, we have
\begin{equation}
Y_{T}(s) =2\langle E_{0}(s)|\rho_{T}(s)|E_{0}(s)\rangle -1 =
\bm{v}_{T}(s)\cdot\bm{q}(s),
\end{equation}
where $\bm{v}_{T}$ is the Bloch vector representing
$\rho_{T}$. Similarly $Y_{0}(s)=\bm{v}_{0}(s)\cdot\bm{q}(s)$, where
$\bm{v}_{0}(s)$ is the Bloch vector corresponding to $\rho_{0}(s)$
which is the solution of Eq.~(\ref{widoppen}) with $T=0$, and initial
condition $\rho_{0}(0) = |E_{0}(0)\rangle\langle E_{0}(0)|$. The
solution is $\rho_{0}(s) = |E_{0}(0)\rangle\langle E_{0}(0)|$.  We may
conclude that $\bm{v}_{0}(s) =
\bm{q}(0)$. Hence, $Y_{T}(s)\geq Y_{0}(s)$ is only 
 Eq.~(\ref{basolikhet}) in disguise.  $\Box$\newline

If we combine Eq.~(\ref{tvaanivaa}) with Eq.~(\ref{Yineq}) and let $A
= 0$, $B = 1$, 
we find the following inequality
\begin{equation}
\label{grdlolik}
1- Y_{T}(1) \geq I_{0},
\end{equation}
where
\begin{equation}
\label{Iintegdef}
I_{0} = 4 \int_{0}^{1}\int_{0}^{s'}
e^{-T[Q(s')-Q(s'')]}Z(s')Z(s'')Y_{0}(s'')ds''ds'
\end{equation}
and where we have used that $Y_{T}(0) = 1$.

This implies that if $Y_{T}(1)$ is to approach $1$, then $I_{0}$ has
to go to zero, or be negative, since $Y_{T}\leq 1$. However,
$I_{0}\geq 0$, as is shown below.

Equation (\ref{Ynoll}) yields
\begin{equation}
\label{Ynolikhet}
Y_{0}(s)= \bm{q}(s)\cdot\bm{q}(0) \geq
\frac{-(N-1)(2s-1)}{\sqrt{N}\sqrt{1 + (N-1)(2s-1)^{2}}}.
\end{equation}
By inserting Eq.~(\ref{Ynolikhet}) into Eq.~(\ref{Iintegdef}) and by
making the change of variables $x = \sqrt{N-1}(2s'-1)$ and $y =
\sqrt{N-1}(2s''-1)$ one obtains
\begin{equation}
\label{mvbdiu}
I_{0} \geq
\sqrt{\frac{N-1}{N}}\int_{-\sqrt{N-1}}^{\sqrt{N-1}}\int_{-\sqrt{N-1}}^{x}
\frac{ -y e^{-T[Q(\kappa(x))-Q(\kappa(y))]}}{(1+x^{2})(1+y^{2})^{3/2}}dydx.
\end{equation}
In the above expression we have introduced the function $\kappa(x) = 1/2
+x/(2\sqrt{N-1})$.  Furthermore,
\begin{eqnarray}
\label{Qdefeni}
TQ\bm{(}\kappa(x)\bm{)} &=&
T\int_{0}^{\kappa(x)}\frac{1}{N^{\sigma}}[1+(N-1)(2s'-1)^{2}]^{\sigma}ds'
\nonumber\\& = & \alpha [\Phi(x)-\Phi(-\sqrt{N-1})],
\end{eqnarray}
where
\begin{equation}
\label{defalphaglobal}
\alpha = \frac{T}{2N^{\sigma}\sqrt{N-1}}
\end{equation}
and 
\begin{equation}
\Phi(x) = \int_{0}^{x}(1+{x'}^{2})^{\sigma}dx'.
\end{equation}
Note that $\Phi(-x) = -\Phi(x)$.
By using Eq.~(\ref{Qdefeni}) one can write
\begin{equation}
TQ\bm{(}\kappa(x)\bm{)}-TQ\bm{(}\kappa(y)\bm{)}
 = \alpha\left[\Phi(x)-\Phi(y)\right].
\end{equation}
Define 
\begin{equation}
\label{defIab}
I(\alpha,\beta) = \int_{-\beta}^{\beta}f_{\alpha}(x)\int_{-\beta}^{x}
g_{\alpha}(y)dydx,
\end{equation}
\begin{equation}
f_{\alpha}(x) = e^{-\alpha \Phi(x)}\frac{1}{1+x^{2}},\quad
g_{\alpha}(y) = e^{\alpha \Phi(y)} \frac{ -y
}{(1+y^{2})^{3/2}}\nonumber.
\end{equation}
From Eq.~(\ref{mvbdiu}) it follows that $I_{0}$ fulfills the
 inequality
\begin{equation}
\label{IIineq}
I_{0} \geq \sqrt{\frac{N-1}{N}}I(\alpha,\sqrt{N-1}).
\end{equation}
The definition in Eq.~(\ref{defIab}) yields
\begin{eqnarray}
\label{dIdb}
\frac{d}{d\beta}I(\alpha,\beta) & = & 
f_{\alpha}(\beta)\int_{-\beta}^{\beta} g_{\alpha}(y)dy +
g_{\alpha}(-\beta)\int_{-\beta}^{\beta}f_{\alpha}(x)dx\nonumber\\ &= &
- 2\frac{e^{-\alpha \Phi(\beta)}}{1+\beta^{2}}\int_{0}^{\beta}
\frac{y\sinh\bm{(}\alpha \Phi(y)\bm{)}
}{(1+y^{2})^{3/2}}dy\nonumber\\ & & + 2\frac{\beta e^{-\alpha \Phi(\beta)}
}{(1+\beta^{2})^{3/2}}\int_{0}^{\beta}\frac{\cosh\bm{(}\alpha
\Phi(x)\bm{)}}{1+x^{2}}dx.\nonumber\\
\end{eqnarray}
The second equality in the expression above is obtained by separating
the integrals ``$\int_{-\beta}^{\beta}$'' into ``$\int_{-\beta}^{0}
+\int_{0}^{\beta} $'' and make the change of variables $x\rightarrow
-x$ (and $y\rightarrow -y$) in the $\int_{-\beta}^{0}$ integrals.

If one uses the inequality
\begin{equation}
\label{ineqspec}
\frac{y}{(1+y^{2})^{3/2}}\leq
 \frac{\beta}{\sqrt{1+\beta^{2}}}\frac{1}{1+y^{2}}, \quad \forall y\in
 [0,\beta].
\end{equation}
in Eq.~(\ref{dIdb}), one obtains 
\begin{equation}
\label{Fdefineq}
\frac{d}{d\beta}I(\alpha,\beta)  
\geq  2\frac{\beta e^{-\alpha \Phi(\beta)} }{(1+\beta^{2})^{3/2}}
\int_{0}^{\beta} \frac{e^{-\alpha \Phi(x)}}{1+x^{2}}dx \equiv F(\alpha,\beta).
\end{equation}
Since the integrand of the above integral is strictly positive, one
can conclude that
\begin{equation}
\frac{d}{d\beta}I(\alpha,\beta)  > 0,\quad \beta > 0.
\end{equation}
Hence, for a fixed $\alpha$ the function $I(\alpha,\beta)$ is strictly
increasing. Similarly, $\sqrt{(N-1)/N}$ increases with $N$. If we
assume $N\geq 2$ it follows that
\begin{eqnarray}
\label{samlat}
I_{0} &\geq & \frac{1}{\sqrt{2}}I(\alpha,\sqrt{N-1}) \geq
\frac{1}{\sqrt{2}}I(\alpha,1)\nonumber \\
&\geq &
\frac{1}{\sqrt{2}}\int_{0}^{1}F(\alpha,\beta')d\beta' > 0.
\end{eqnarray}
We can conclude that if $I_{0}\rightarrow 0$,
then $I(\alpha,1)\rightarrow 0$, necessarily. This, in turn, implies
$\int_{0}^{1}F(\alpha,\beta')d\beta'\rightarrow 0$.  We now
investigate how the last expression depends on $\alpha$.
\begin{equation}
\frac{d}{d\alpha}\int_{0}^{1}F(\alpha,\beta')d\beta' 
= \int_{0}^{1}\frac{d}{d\alpha}F(\alpha,\beta')d\beta',
\end{equation}
\begin{eqnarray}
\label{sistaglobal}
\frac{dF}{d\alpha} & = &
 - 2\frac{\beta \Phi(\beta) e^{-\alpha \Phi(\beta)} }{(1+\beta^{2})^{3/2}}
\int_{0}^{\beta} \frac{e^{-\alpha \Phi(x)}}{1+x^{2}}dx\nonumber\\
& & -2\frac{\beta e^{-\alpha \Phi(\beta)}
}{(1+\beta^{2})^{3/2}}\int_{0}^{\beta}
\frac{\Phi(x)e^{-\alpha
\Phi(x)}}{1+x^{2}}dx.\nonumber\\
\end{eqnarray}
Both terms on the right-hand side in the above expression are
strictly less than zero, for $\beta >0$.  It follows that
$\int_{0}^{1}F(\alpha,\beta')d\beta'$ is a strictly 
decreasing function of $\alpha$.

Thus, a necessary condition for the integral to go to zero is that
$\alpha$ goes to infinity.  Due to the form of $\alpha$ in
Eq.~(\ref{defalphaglobal}) it follows, for large $N$, that
$TN^{-(\sigma+1/2)}\rightarrow \infty$ is a necessary condition for
$Y_{T}(1)\rightarrow 1$. Since $Y_{T}(1)\rightarrow 1$ is equivalent
to $p\rightarrow 1$, we have shown that the sufficient condition $T\gg
N^{\sigma +1/2}$, proved in Sec.~\ref{sec:globsuff}, is also
necessary.

\subsubsection{\label{sec:locnecess}Local search}
The proof of the necessary condition in the local case resembles
rather closely the proof of the global search case.  However, we
assume that $\sigma \geq 1$. We let $\widetilde{Y}_{T}$ denote the
solution of Eq.~(\ref{tvaanivaa}) for a given run time $T$, with $A=0$
and $B=1$, and with $Q$ replaced by $\widetilde{Q}$ defined in
Eq.~(\ref{wQWH}). Similarly we let $\widetilde{\rho}_{T}$ denote the
solution of Eq.~(\ref{widoppenlocal}), for the given run time $T$.
 
It is straightforward to prove the counterpart of Lemma \ref{lemmatre}
\begin{equation}
\label{Yineqlocal}
\widetilde{Y}_{T}(r)\geq \widetilde{Y}_{0}(r),
\quad T\geq 0,\quad 0\leq r\leq 1.
\end{equation}
Similarly we can obtain $1-\widetilde{Y}_{T}(1)\geq \widetilde{I}_{0}$
with $\widetilde{I}_{0}$ as in Eq.~(\ref{Iintegdef}), but with $Q$
replaced by $\widetilde{Q}$. Moreover,
\begin{equation}
\label{Ynolikhetlocal}
\widetilde{Y}_{0}(r)= \bm{q}(r)\cdot\bm{q}(0) \geq
\frac{-(N-1)(2r-1)}{\sqrt{N}\sqrt{1 + (N-1)(2r-1)^{2}}}.
\end{equation}
Due to Eq.~(\ref{Ynolikhetlocal}), Eq.~(\ref{mvbdiu}) remains true if
we replace $I_{0}$ with $\widetilde{I}_{0}$ and $Q(\kappa(x))$ with
$\widetilde{Q}(\kappa(x))$. We also obtain
\begin{equation}
\label{Qdefeniloc}
T\widetilde{Q}\bm{(}\kappa(x)\bm{)}
 =  \widetilde{\alpha} [\widetilde{\Phi}(x)-\widetilde{\Phi}(-\sqrt{N-1})],  
\end{equation}
where
\begin{equation}
\label{defalphalokal}
\widetilde{\alpha} = \frac{T}{2LN^{\sigma-1}\sqrt{N-1}} 
= \frac{T}{2N^{\sigma}\arctan(\sqrt{N-1})}
\end{equation}
and
\begin{equation}
\widetilde{\Phi}(x) = \int_{0}^{x}(1+{x'}^{2})^{\sigma-1}dx'.
\end{equation}

Now we define $\widetilde{I}(\widetilde{\alpha},\beta)$ as in
Eq.~(\ref{defIab}) but with $\alpha$ replaced by $\widetilde{\alpha}$,
and $f_{\alpha}$, $g_{\alpha}$ replaced by
\begin{equation}
\widetilde{f}_{\widetilde{\alpha}}(x) = 
e^{-\widetilde{\alpha}
\widetilde{\Phi}(x)}\frac{1}{1+x^{2}},
\quad \widetilde{g}_{\widetilde{\alpha}}(y) =
e^{\widetilde{\alpha}\widetilde{\Phi}(y)} 
\frac{-y}{(1+y^{2})^{3/2}}.
\end{equation}
The line of reasoning from Eq.~(\ref{IIineq}) to
Eq.~(\ref{sistaglobal}) is unaltered up to a replacement of
$s,\alpha,\Phi,I_{0}$, and $I(\alpha,\beta)$ with
$r,\widetilde{\alpha},\widetilde{\Phi},\widetilde{I}_{0}$, and
$\widetilde{I}(\widetilde{\alpha},\beta)$. We define
$\widetilde{F}(\widetilde{\alpha},\beta)$ in analogy with
$F(\alpha,\beta)$ in Eq.~(\ref{Fdefineq}).

Analogously to the global search case we find that a necessary
condition for the limit $\widetilde{Y}_{T}(1)\rightarrow 1$ is that
$\widetilde{\alpha}\rightarrow \infty$, since
$\int_{0}^{1}\widetilde{F}(\widetilde{\alpha},\beta')d\beta'$ is a
strictly decreasing function in $\widetilde{\alpha}$.  For large $N$,
it follows, due to the form of $\widetilde{\alpha}$ in
Eq.~(\ref{defalphalokal}) that $TN^{-\sigma}\rightarrow \infty$ is a
necessary condition for $p\rightarrow 1$. Thus, we have proved that
the sufficient condition for adiabaticity, proved in
Sec.~\ref{sec:locsuff}, is also necessary.

\section{\label{sec:success}Success probabilities and the 
$p\rightarrow 1$ limit} We have so far investigated the conditions for
the success probability $p$ to approach unity, which echoes the
formulation of the standard adiabatic theorem. However, for an
actual implementation of the adiabatic quantum computer we have to be
satisfied with a success probability less than unity in order to have
a finite run time. It thus seems reasonable to ask how much of the
results obtained for the $p\rightarrow 1$ limit survives when $p<1$.

Consider the set of pairs of list lengths and run times $(N,T)$ that
gives precisely the success probability $p$. One may note that $T$
does not necessarily have to be a function of $N$, as there may
accidentally be several values of $T$, for a fixed $N$, which give the
same $p$.  We show that the run time $T$ is bounded by curves with the
same asymptotic behavior as found in the $p\rightarrow 1$ cases.

We assume that the system initially is in the ground state,
i.e., $\rho_{00}(0) =1$ and $\rho_{01}(0) = 0$. We denote $p =
\rho_{00}(1)$. The strategy is to obtain expressions of the form
$1-p\leq U(T/N^{\nu})$ or $1-p\geq V(T/N^{\nu})$. If $U$ and $V$ are
continuous and strictly decreasing functions, we may invert them, on
suitable domains, and obtain bounds $N^{\nu}V^{-1}(1-p)\leq T\leq
N^{\nu}U^{-1}(1-p)$. We derive both upper and lower bounds in the
wide-open case.  In the semi-open case we obtain upper bounds only.

One may note that the upper bounds can be seen as sufficient
conditions on the run time to achieve a success probability which is
at least $p$.

\subsection{The wide-open case, global and local}
An upper bound for the global wide-open case can be obtained
 directly from Eq.~(\ref{globwo}). The result is
\begin{equation}
\label{wogluppett}
T \leq N^{\sigma+1/2}\frac{\pi}{1-p},\quad \sigma\geq 0.
\end{equation}
We obtain a lower bound by using results from
Sec.~\ref{sec:globnecess}.  By combining Eqs.~(\ref{grdlolik}) and
(\ref{samlat}) one obtains
\begin{equation}
1-p \geq \frac{1}{2\sqrt{2}}\int_{0}^{1}F(\alpha,\beta)d\beta \equiv
C(\alpha),
\end{equation}
where $F$ is defined in Eq.~(\ref{Fdefineq}), and $\alpha$ in
Eq.~(\ref{defalphaglobal}).  We know from Sec.~\ref{sec:globnecess}
that $C$ is strictly decreasing in $\alpha$. Moreover, $C$ is
continuous. Hence, $C$ is invertible and
\begin{equation}
\label{wogllow}
T \geq 2N^{\sigma}\sqrt{N-1} C^{-1}(1-p),\quad N\geq 2.
\end{equation}
The bounds in Eqs.~(\ref{wogluppett}) and
(\ref{wogllow}) fulfill the same $N^{\sigma+1/2}$ rate as we found in the
$p\rightarrow 1$ investigation.

For the wide-open local case we may use Eq.~(\ref{wioloc}), and $p=
\rho_{00}(1) = \widetilde{\rho}_{00}(1)$ to obtain
\begin{equation}
T\leq \frac{\pi^{2}}{2}\frac{N^{\sigma}}{1-p}.
\end{equation}
Analogously to the global search case we can obtain a lower bound
to the wide-open local search from Sec.~\ref{sec:locnecess}, as
\begin{equation}
1-p \geq \frac{1}{2\sqrt{2}}
\int_{0}^{1}\widetilde{F}(\widetilde{\alpha},\beta)d\beta \equiv
\widetilde{C}(\widetilde{\alpha}),
\end{equation}
with $\widetilde{\alpha}$ as in
 Eq.~(\ref{defalphalokal}). This results in the bound
\begin{equation}
T\geq   2N^{\sigma}\arctan(\sqrt{N-1}) \widetilde{C}^{-1}(1-p),
\end{equation}
which, for large $N$, increases as $N^{\sigma}$.

We have found that for a fixed success probability $p$ the run time is
locked between two bounds. These bounds have the same scaling with the
list length $N$ as found in the $p\rightarrow 1$ limit.

\subsection{The semi-open case, global and local}
In the semi-open global search case we use Eq.~(\ref{sogl}), with
$\rho_{01}(0) = 0$ and $\rho_{00}(0) = 1$ to obtain
\begin{equation}
\label{boundsoglob}
T \leq \pi N\frac{BK+6A}{A^{2}(1-p)}.
\end{equation}
In the semi-open local search case one may use Eq.~(\ref{soloc}) to obtain
\begin{equation}
\label{boundsoloc}
T \leq \frac{\sqrt{N}}{1-p}\frac{2\pi^{2}}{A}\left(1 +
\frac{\sqrt{2}BK}{4A} \right), \quad N\geq 2.
\end{equation}
The bounds in Eqs.~(\ref{boundsoglob}) and (\ref{boundsoloc}) are sufficient
conditions on the run time to obtain at least the success
probability $p$, for a given list length $N$ \cite{sideremark}. 

\subsection{Numerical analysis}

\begin{figure}[h]
\includegraphics[width = 8.5cm]{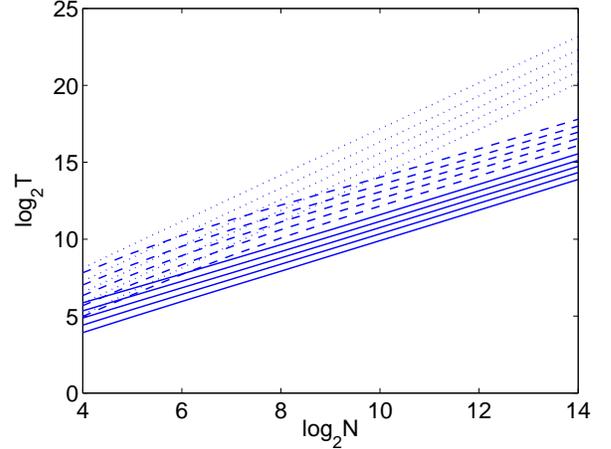}
\caption{\label{fig:globalpvar}
(Color online) Global search with $W(s) = H(s)$. Each curve shows $\log_{2}T$
vs $\log_{2}N$, where $T$ is the run time needed to obtain a given
success probability $p$ at $s = 1$, and where $N$ is the list
length. The dimensionless run time $T$ is the quotient between the
physical run time and $T_{0}$ defined in Eq.~(\ref{defT0}).  The
dotted curves correspond to the wide-open case, defined by the degree
of openness $\omega =1$. The dashed curves correspond to the semi-open
case $\omega = 0.9$, and the solid correspond to $\omega = 0.5$.
Within each group, each curve corresponds to a success probability,
which counted from below is $p =
0.4,0.5,0.6,0.7,0.8$. As seen, the asymptotic slope appears to be
independent of the choice of success probability $p$. For the
wide-open case the asymptotic slope seems to tend to $\nu = 3/2$,
independent of $p$. For the semi-open cases the asymptotic slope
appears to tend to $\nu = 1$, independent of $p$.}
\end{figure}

We have numerically calculated the actual run times of the wide-open
global and local search. Let us choose $A = \cos(\omega\pi/2)$ and
$B = \sin(\omega\pi/2)$. This choice of $A$ and $B$ is merely a
convenient way to obtain an interpolation between the closed case
$\omega = 0$ and the wide-open case $\omega = 1$, without effectively
changing $T$. Moreover we have chosen $W(s) = H(s)$. Thus, we consider
an environment that monitors the energy of the system. The initial
condition at $s = 0$ is $\rho(0)= |\psi\rangle\langle\psi|$ with
$|\psi\rangle$ as in Eq.~(\ref{eqalsuperpos}). In other words, we
begin at the initial ground state of the Hamiltonian in
Eq.~(\ref{family}).

We should also comment upon the units. As seen in
Sec.~\ref{sec:search} we have put the energy gap between the ground
state and the degenerate first excited state to $1$ in the initial
Hamiltonian $H(0)$. Depending on the actual system used to realize the
search procedure, the energy gap is actually $\Delta\mathcal{E}$ in
some appropriate energy unit. To put the energy gap to $1$ corresponds
to measuring energy in units of $\Delta\mathcal{E}$. Furthermore, as
we have chosen $\hbar = 1$, it follows that the run time $T$, used
throughout this paper, can be seen as the quotient $T =
\overline{T}/T_{0}$, of the physical run time $\overline{T}$ and
$T_{0}$ defined by
\begin{equation}
\label{defT0}
T_{0} = \Delta\mathcal{E}^{-1}.
\end{equation}

\begin{figure}[h]
\includegraphics[width = 8.5cm]{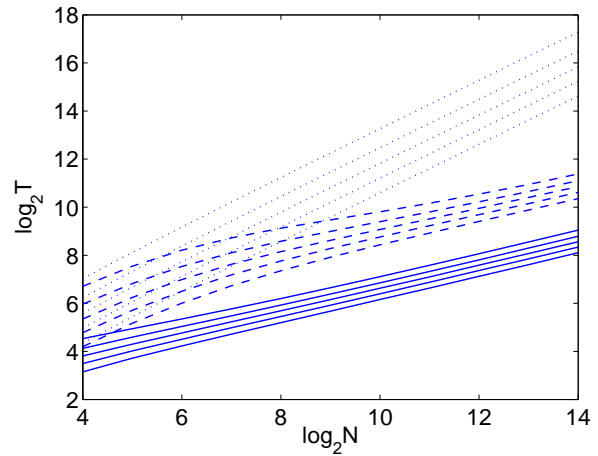}
\caption{\label{fig:localpvar} (Color online) Local search with $W(s) = H(s)$. 
Each curve shows $\log_{2}T$ vs $\log_{2}N$, where $T$ is the run
time needed to obtain a given success probability $p$ at $s = 1$, and
where $N$ is the list length.  The dimensionless run time $T$ is the
quotient between the physical run time and $T_{0}$ defined in
Eq.~(\ref{defT0}). Similar to Fig.~\ref{fig:globalpvar} the dotted
curves correspond to the wide-open case, defined by the degree of
openness $\omega = 1$. The dashed curves correspond to the semi-open
case $\omega = 0.9$, and the solid curves correspond to $\omega =
0.5$.  Within each group each curve corresponds to a success
probability $p = 0.4,0.5,0.6,0.7,0.8$, counted from below. Similarly
to the global search, the asymptotic slope appears to be independent
of the choice of success probability $p$. For the wide-open case the
asymptotic slope seems to tend to $\nu = 1$, independent of $p$. For
the semi-open cases the asymptotic slope appears to tend to the $\nu =
1/2$, independent of $p$.}
\end{figure}

In Fig.~\ref{fig:globalpvar}, the dotted curves correspond to the
wide-open global search with success probability ranging from $p =
0.4$ to $p = 0.8$. The curves seem to follow closely to a linear
asymptote with slope $\nu = 3/2$, independent of the choice of
$p<1$. The dotted curves in Fig.~\ref{fig:localpvar} similarly
correspond the wide-open local search case. The asymptotic slope of
these curves seems to be $\nu = 1$ for every choice of $p<1$.

We have also calculated the actual run times of the semi-open global
and local search numerically. Again we use $W(s) = H(s)$, $A =
\cos(\omega\pi/2)$, and $B =
\sin(\omega\pi/2)$. The results for the global case 
are shown in Fig.~\ref{fig:globalpvar}, in form of the dashed curves
corresponding to $\omega = 0.9$, and the solid curves corresponding to
$\omega = 0.5$. All these curves seem to tend to the asymptotic slope
$\nu = 1$, which is in line with our analytical results.  In
Fig.~\ref{fig:localpvar} we similarly have created plots for the
semi-open local search case. These again correspond to the dashed
curves with $\omega = 0.9$, and the solid curves with $\omega =
0.5$. The results are qualitatively similar to the semi-open global
search case, but with asymptotic slope $\nu = 1/2$.

\begin{figure}[h]
\includegraphics[width = 8.5cm]{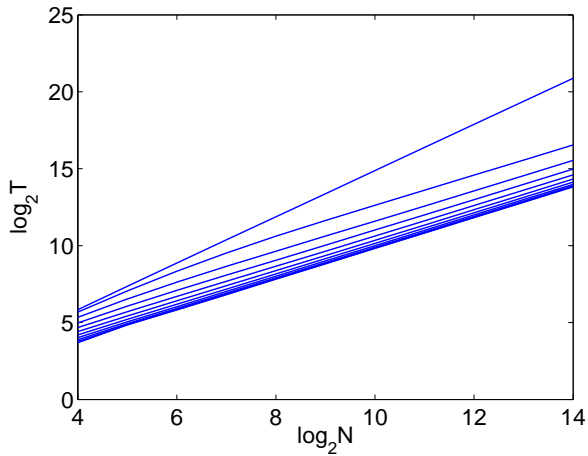}
\caption{\label{fig:globalw} (Color online) Global search with
$W(s) = H(s)$ and success probability $p = 1/2$. The curves show
$\log_{2}T$ vs $\log_{2}N$, where $T$ is the run time needed to obtain
the success probability $p = 1/2$ at $s = 1$, and where $N$ is the
list length.  The dimensionless run time $T$ is the quotient between
the physical run time and $T_{0}$ defined in Eq.~(\ref{defT0}).  Each
line shows the result for a given value of the degree of openness
$\omega$. Counted from below, the lines correspond to $\omega =
0,0.1,\ldots,0.9,1$. Hence, the uppermost line corresponds to the
wide-open case $\omega = 1$ and the lowermost line corresponds to the
closed case $\omega = 0$. All the lines seem to tend to the slope $\nu
= 1$, except the uppermost which seems to tend to $\nu = 3/2$.}
\end{figure}

\begin{figure}[h]
\includegraphics[width = 8.5cm]{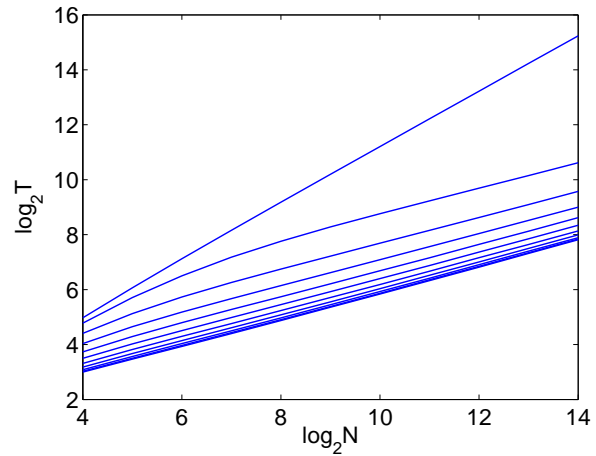}
\caption{\label{fig:localw} (Color online) Local search with 
$W(r) = H(r)$ and success
probability $p = 1/2$. The curves show $\log_{2}T$ vs $\log_{2}N$
plots, where $T$ is the run time needed to obtain the success
probability $p = 1/2$ at $s = 1$, and where $N$ is the list length.
The dimensionless run time $T$ is the quotient between the physical
run time and $T_{0}$ defined in Eq.~(\ref{defT0}).  Each line shows
the result for a given value of the degree of openness
$\omega$. Counted from below, the lines correspond to
$\omega = 0,0.1,\ldots,0.9,1$. All the lines seem to tend to the slope
$\nu = 1/2$, except the uppermost which appears to tend to $\nu = 1$.}
\end{figure}

Figures \ref{fig:globalw} and \ref{fig:localw} show the global and
local search case for a fixed success probability $p = 1/2$ but for
different degrees of openness $\omega$.  Figure \ref{fig:globalw}
suggests that in the semi-open cases the asymptotic slope is $\nu =
1$, while in the wide-open case the asymptotic slope is $\nu = 3/2$,
which is consistent with our analytical results.  In
Fig.~\ref{fig:localw} the curves in the local search case appears to
have a qualitatively similar behavior to the global search case,
except that the asymptotic slope is $\nu = 1$ in the wide-open case,
and $\nu = 1/2$ in the semi-open case.

From Figs.~\ref{fig:globalw} and \ref{fig:localw} one may obtain an
intuitive understanding of the ``discontinuity'' of the scaling of the
run times, when going from the semi-open to the wide-open case.  The
closer we get to the wide-open case (i.e., the closer $\omega$ gets to
$1$), the higher up in list-lengths the curve ``follows'' the curve of
the wide-open case, before it turns to its asymptotic slope.  In
this context one should also note that for every fixed list-length
$N$, the system is continuous with respect to $\omega$. In other
words, the discontinuity only appears for the asymptotes in the
$N\rightarrow\infty $ limit, not in any system corresponding to a
fixed list. The list-length $N$ corresponds to the dimension of
the total Hilbert space, but in the effective two-dimensional Hilbert
space the list-length only enters as a parameter in the Hamiltonian.
If we are to understand the discontinuity of the asymptotic scaling as
a function of the degree of openness, in the case $W(s) = H(s)$, we
may consider how the parameter $N$ enters the terms $A[H,\rho]$ and
$B[H,[H,\rho]]$ in the master equation. The minimum energy gap of the
Hamiltonian is proportional to $1/\sqrt{N}$. Thus, at the minimum gap
the list length $N$ enters as $A/\sqrt{N}$ in the Hamiltonian part of
the master equation, but as $B/N$ in the decohering part.  No matter
how small $A$ is, as long as it is nonzero, $A/\sqrt{N}$ always
dominates $B/N$ in the limit of large $N$. Although this, strictly
speaking, only holds at $s=1/2$, it seems intuitively reasonable that
the Hamiltonian term dominates the evolution, for large $N$, as long
as $A\neq0$.  Intuitively it also seems reasonable that if the
Hamiltonian dominates the evolution, then the Hamiltonian
determines the asymptotic scaling of the run time, and thus the
scaling becomes the one of the ideal closed case.  However, if $A=0$
then the decohering term determines the scaling.

\section{\label{sec:concl}Conclusions}
We investigate the effect of decoherence on the adiabatic search of a
marked element in a disordered list of $N$ elements. More specifically
we consider decoherence with respect to the instantaneous eigenbasis
of the adiabatic search Hamiltonian. The decoherence is modeled by
using a master equation of the form $\dot{\rho} = -iA[H,\rho]
-B\bm{[}W,[W,\rho]\bm{]}$ (with $\hbar =1$), where $H$ is the time-dependent
search Hamiltonian and $W$ is a time-dependent Hermitian operator that
commutes with $H$. The operator $W$ is only subjected to the condition
in Eq.~(\ref{condition}), which, in some sense, states that the fluctuations
of $W$ (with respect to time) should not grow with the list-length
$N$.

We prove sufficient conditions for the $p\rightarrow 1$ limit of the
success probability $p$. In the semi-open case ($A\neq 0$) we prove
that these sufficient conditions are $T\gg N$ and $T \gg \sqrt{N}$ in
the global and local search case, respectively, where $T$ is the run time
of the adiabatic search. In other words, the results for the ideal
adiabatic search ($B = 0$) remains valid even in the presence of
decoherence with respect to the instantaneous eigenbasis.

In the wide-open case ($A = 0$), sufficient and necessary conditions
for the $p\rightarrow 1$ limit, are deduced. In the case where $W = H$
we find that the conditions for $p\rightarrow 1$ are $T\gg N^{3/2}$
and $T \gg N$ for the global and local search, respectively.

We finally investigate how the run time $T$ depends on $N$ for a fixed
success probability $p$ less than unity. We find that it is sufficient
with run times that scales like $N$ and $\sqrt{N}$, in the semi-open
case, as for the ideal global and local adiabatic searches. In the
wide-open case we instead find the scalings $N^{3/2}$ and $N$, for the
global and local search, respectively.
  
In view of these results the following picture emerges. In the
semi-open cases the Hamiltonian dynamics dominates the behavior of the
system for sufficiently large list lengths. To be more precise, an
increased degree of eigenbasis decoherence does increase the run time
of the adiabatic search, but only through the constant $C$ of the
asymptote $CN^{\nu}$.  In the wide-open case, however, the protective
effect of the Hamiltonian dynamics is not present. Consequently, the
asymptotic behavior depends directly on the choice of $W$. As a
consequence the asymptotic behavior of the wide-open case may be
distinctly different from that of the semi-open case, although we have
the same choice of $W$. One may note that the abrupt change in
asymptotic behavior with respect to the transition from semi-open to
wide-open system, raises a warning sign concerning the use of the
wide-open evolution as an approximation to an ``almost'' wide-open
evolution. It should be noted, however, that for any fixed list
length, the system is continuous with respect to the degree of
openness, and that the discontinuity rather concerns the asymptotic
scaling of the run time in the limit of large list-lengths.

The decoherence model provided by Eq.~(\ref{enkel}) implies that we
disregard the memory effects in the environment and the corresponding
fluctuations, which seems justifiable if the relevant time scales of
the evolution is much larger than the dominating time scales of the
memory of the environment. The modeling of memory effects generally
leads to integro-differential equations \cite{Nakajima,Zwanzig}, or to
time-local non-Markovian equations
\cite{Breuer}. To include such effects
into the analysis of the influence of the environment on the
efficiency of the adiabatic quantum computer would be interesting but
challenging.

Finally we note that, in spite of the apparent difference between the
adiabatic quantum computing scheme and the traditional circuit model,
it has been shown that these two models are, in a certain sense,
equivalent
\cite{Aharonov04,Siu}. However, this equivalence does not concern the
robustness to noise, relaxation, or decoherence. Previous findings
suggest that the adiabatic quantum computer to a certain extent should
be resilient against various kinds of open system effects. The results
of this paper provide further evidence for this.


\begin{thebibliography}{99}
\bibitem{note} If we consider Hamiltonians with
a time-independent eigenbasis, then randomness in the eigenenergies
causes decoherence with respect to the eigenstates, when averaged over
several instances. Although this gives rise to decoherence in the
eigenbasis, it is not necessarily of a form that can be expressed as
in Eq.~(\ref{enkel}); this depends on the statistics of the
fluctuations. A time-independent two-level system with an energy gap
that is drawn from an ensemble with Lorentzian
distribution, results in an averaged evolution that can be described
by Eq.~(\ref{enkel}) with time-independent $H$ and $W$. In
Eq.~(\ref{enkel}) the eigenbasis of the Hamiltonian depends on
time. However, in the adiabatic limit many aspects of the above
time-independent case seems likely to be retained.
\bibitem{error}
Note that the quotient $KB/A^{2}$ in Eq.~17 in \cite{aberg05} should be
replaced by $KB/(4A^{2})$.
\bibitem{sideremark}
Note that Eq.~(\ref{boundsoloc}) with $B=0$ gives an alternative proof
of the $\sqrt{N}$ efficiency of the local search algorithm
\cite{Roland02,vanDam01}.
\end{thebibliography}
\end{document}